\documentclass[aip,pof,reprint,amsmath,amssymb]{revtex4-2}
\usepackage{graphicx}
\usepackage{placeins}
\usepackage[letterpaper,hmargin=1in,vmargin=0.95in]{geometry}
\usepackage{booktabs}
\usepackage{xcolor}
\usepackage[colorlinks=true,linkcolor=blue,citecolor=blue,urlcolor=blue]{hyperref}
\hypersetup{pdftitle={Gas radiation and the gravity limit of flicker in a co-flowing hydrogen jet diffusion flame},pdfauthor={Tianfang Xie and Meng Wang}}
\newcommand{\Uco}{U_\mathrm{co}}
\newcommand{\Frco}{\mathrm{Fr}_\mathrm{co}}
\newcommand{\alttext}[1]{}   % SI alt text (AIP) not shown in the preprint
\newcommand{\RR}{\raggedright\let\\\tabularnewline}
\newcommand{\figfile}[2][width=\textwidth]{\IfFileExists{#2}{\includegraphics[#1]{#2}}%
  {\fbox{\parbox[c][3cm][c]{0.9\linewidth}{\centering\small Figure in preparation: \texttt{\detokenize{#2}}}}}}

\begin{document}

\title{Gas radiation and the gravity limit of flicker in a co-flowing hydrogen jet diffusion flame}

\author{Tianfang Xie}
\email{tianfangxie@purdue.edu}
\affiliation{School of Aeronautics and Astronautics, Purdue University, West Lafayette, Indiana 47907, USA}
\author{Meng Wang}
\affiliation{School of Automotive Engineering, Hunan Industry Polytechnic, Changsha 410208, China}

\begin{abstract}
Although weak at normal gravity, gas radiation sets the gravity at which a co-flowing hydrogen jet diffusion flame stops flickering. Flamelet-based simulations with unity Lewis numbers, which reproduce the flicker frequency measured by femtosecond coherent anti-Stokes Raman scattering within 4\%, are combined with local stability analysis. At one tenth of normal gravity, the adiabatic flame flickers, the radiating flame does not, and switching the emission off restores flicker. Radiation acts through two mechanisms that do not weaken at reduced gravity. The hot outer layer loses buoyancy in proportion to the gas residence time, and a disturbance that displaces the flame also displaces its emission, which damps the disturbance on the lean flank (displaced-emission damping). Freezing the radiative source, which removes only its response to disturbances, slows the downstream recession of the oscillation more than tenfold. Both mechanisms slow the upstream spreading of disturbances. This spreading, competing with the co-flow, sets the onset location and the critical co-flow. Without radiation, the spreading speed scales with the square root of gravity; its radiative reduction does not. At reduced gravity, the flicker limit therefore becomes a straight line relating the square root of the co-flow Froude number to a radiation number, the ratio of buoyant to radiative time scales. The calibrated line places the critical gravity for the experimental co-flow at 0.13 times normal gravity, 2.6 times the adiabatic value. It correctly predicted whether two further flames keep flickering. Beyond the limit, transient wave packets can mimic flicker throughout a short drop-tower test.
\end{abstract}

\maketitle
{\noindent\footnotesize This work is licensed under a Creative Commons Attribution-NonCommercial 4.0 International License (CC BY-NC 4.0), \url{https://creativecommons.org/licenses/by-nc/4.0/}.\par}
\medskip

%=====================================================================================================================
\section{Introduction}
\label{sec:intro}

Buoyant diffusion flames oscillate at a low frequency, typically 10--20\,Hz for burners of centimeter scale, a
phenomenon known as flame flicker~\cite{BuckmasterPeters1988,CetegenAhmed1993,Durox1997}. The oscillation is carried by toroidal vortices that form outside the flame and travel downstream along its surface;
they arise in the shear layer between the buoyantly accelerated hot gas and the surrounding
air~\cite{Davis1991,KattaRoquemore1993,XiaZhang2018}. Because these vortices are driven by buoyancy, the flicker
frequency depends only weakly on the fuel or the flow rate and scales approximately as $(g/D)^{1/2}$, where $g$ is
the gravitational acceleration and $D$ the burner diameter~\cite{CetegenAhmed1993,XiaZhang2018}. Experiments at reduced and elevated gravity give $f\propto g^{n}$ with $n$ between about 0.4 and 2/3, depending on the burner size
and the Froude number~\cite{Durox1997,Sato2000}. Reducing gravity and adding an air co-flow both suppress the flicker.
In soot-free hydrogen flames, radiation is weak at normal gravity and is usually neglected. At reduced gravity,
however, the gas stays longer in the flame and loses more of its heat to radiation, while the buoyant drive of the
flicker weakens. This paper asks how gas radiation shifts the gravity at which a co-flowing hydrogen jet diffusion
flame stops flickering.

In a spatially developing flow, a convectively unstable region amplifies disturbances while they are swept
downstream, and the flow returns to its steady state once they have passed. In an absolutely unstable region,
disturbances also spread upstream, and a sufficiently large region of this kind sustains a global mode that oscillates
at its own frequency~\cite{HuerreMonkewitz1990,Chomaz2005}. Light and hot jets become absolutely unstable below a
critical ratio of jet to ambient density~\cite{Monkewitz1990,Lesshafft2006,Coenen2017}. Once a co-flow exceeds a critical fraction of the jet velocity, below about 20\% for helium
jets~\cite{JendoubiStrykowski1994,Hallberg2007}, it sweeps disturbances downstream and removes the absolute
instability, and with it the global oscillation. In jet diffusion flames
without co-flow, linear global analysis~\cite{MorenoBoza2016} and direct simulations of hydrogen--nitrogen jet
flames~\cite{Juniper2009} find the flicker, a global mode, only below a critical Froude number; local analyses overpredict this threshold by more
than a factor of two~\cite{MorenoBoza2016}. Local analyses of methane jet flames place
the absolute instability in the outer, buoyancy-driven shear layer, beginning either where the oscillation starts, 1--1.5 diameters downstream depending on the
co-flow~\cite{Lingens1996,LingensReeker1996}, or at the nozzle exit~\cite{SeeIhme2014}.

In jet flames, an air co-flow raises the flicker frequency, reduces its amplitude, and moves the oscillation
downstream~\cite{LingensReeker1996}; at large co-flow velocities, the outer vortices form beyond the visible flame, and
above a critical ratio of co-flow to jet velocity, the flame surface no longer flickers~\cite{GohariDarabkhani2011}.
For methane flames at normal gravity, this critical ratio correlates with the jet Froude number~\cite{Fujisawa2016},
and a global analysis of buoyant plumes relates it to the Froude number and the density ratio~\cite{BharadwajDas2021}.
A flicker suppressed by the co-flow can be re-excited by acoustic forcing~\cite{PoF2026}. Reducing gravity suppresses the
flicker as well: laminar jet diffusion flames do not flicker in microgravity~\cite{Bahadori1990,Lin1999}, and helium
jets lose their global oscillation in a drop tower~\cite{Yep2003}. The gravity at which the flicker ceases is less
well established. Simulations of hydrogen--nitrogen jet flames without radiation found the flicker still present at
0.05 times normal gravity~\cite{RoquemoreKatta2000}, whereas in an earlier simulation the outer structures were barely
identifiable at half of normal gravity~\cite{EllzeyOran1991}. For a methane flame on a cup burner, simulations predict a steady flame at 0.5 times normal gravity and below;
the authors attribute this cut-off to the disappearance of an absolute instability and expect it near zero gravity for
high-velocity jets from small nozzles~\cite{KattaTakahashiLinteris2004}.

Radiation can suppress the flicker, and its influence grows as gravity decreases. In a sooting ethylene flame at
normal gravity, soot radiation alone suppresses the flicker, the oscillation decaying at an unchanged
frequency~\cite{KattaSoot2009}. In laminar methane flames, reabsorption changes the structure of flickering flames
more than that of steady ones~\cite{WuZhao2020}. In the cup-burner simulations, the radiative heat loss, small at
normal gravity, quenched the tip of the flame at zero gravity~\cite{KattaTakahashiLinteris2004}. Qin et
al.~\cite{Qin2004} simulated partially premixed methane flames brought suddenly from normal to zero gravity: the
flames became nearly steady within 2.2\,s, radiation changed the zero-gravity flame far more than the normal-gravity
one, and the authors noted that a radiation Damk\"ohler number varies inversely with the Froude number. These studies leave open
whether radiation removes the flicker only by cooling the mean flow, and thereby weakening its buoyant drive, or also
by acting on the disturbances themselves, and what then sets the limit beyond which a flame no longer flickers.

Hydrogen flames make it possible to study this question for gas radiation alone. Their radiation comes from water
vapor only: small at normal gravity, it grows with the size of the flame and the residence time of the gas at reduced
gravity, and without soot, it follows the local temperature and composition. In steam-diluted premixed hydrogen
flames, water-vapor radiation also acts through the reactants: the fresh gas absorbs part of the emitted radiation,
which preheats it and advances the onset of thermodiffusive instability~\cite{BenZenou2026}. The reactants of a
hydrogen jet flame do not absorb, so this preheating route is absent. The high diffusivity of hydrogen adds
differential diffusion, whose effect on the onset and the limit of the flicker is examined here with a differential-diffusion flamelet closure~\cite{PitschPeters1998}.

Here, we show that gas radiation sets the gravity at which a co-flowing hydrogen jet diffusion flame stops flickering.
We study the flame whose flicker Dennis et al.~\cite{Dennis2016} measured with femtosecond coherent anti-Stokes Raman
scattering (fs-CARS), combining flamelet-based simulations with local linear stability analysis. Radiation acts on the
flicker through two mechanisms. It reduces the buoyancy of the hot outer layer in proportion to the residence time of
the gas, and it damps the flicker wave: gas that the wave pushes outward across the lean flank of the flame radiates
more than the gas it replaces and loses part of the buoyancy that drives the wave. We call
this second mechanism displaced-emission damping; freezing the radiative source separates it from the first. Both
mechanisms slow the spreading of disturbances upstream through the hot layer, and the competition between this
spreading and the co-flow sets where the flicker begins. Because the spreading speed of the adiabatic flame scales
with the square root of gravity, while its radiative reduction does not, the limit of the flicker at reduced gravity
becomes a straight line relating the square root of the co-flow Froude number to a radiation number. Calibrated on the
simulations, this line predicted the outcome of two flames computed afterwards. Beyond the limit, transient wave
packets can mimic flicker within the few seconds of a drop-tower experiment.

%=====================================================================================================================
\section{Flame configuration and numerical methods}
\label{sec:methods}

\subsection{Reference flame and measurements}
\label{sec:config}
\begin{figure*}[!tbp]
\centering
\figfile{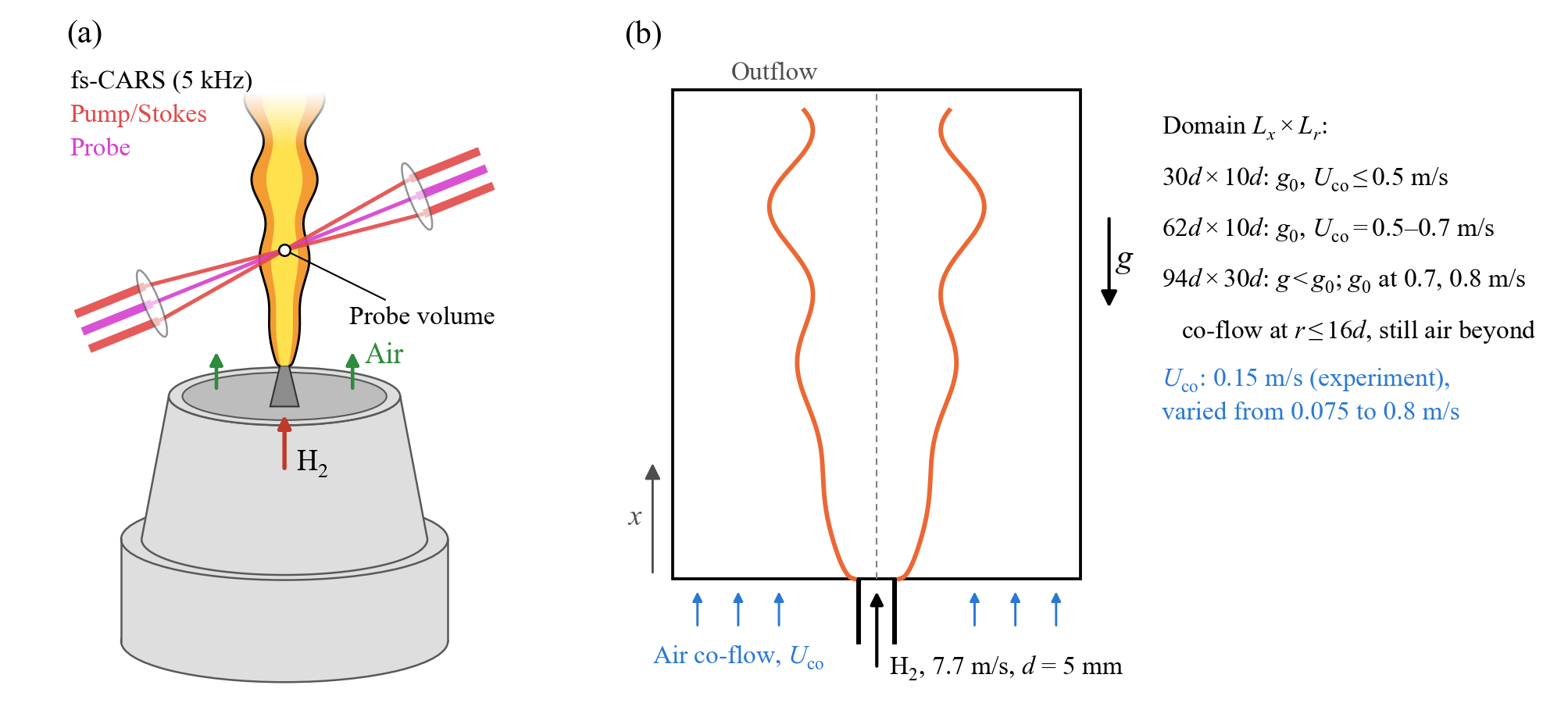}
\caption{(a) Burner of Ref.~\cite{Dennis2016}: hydrogen jet in an air co-flow, with the fs-CARS pump/Stokes and
probe beams crossing at the probe volume (schematic, not to scale). (b) Computational domains (not to scale).}
\label{fig:config}
\end{figure*}

The simulations reproduce the co-flowing hydrogen jet flame of Dennis et al.~\cite{Dennis2016}, which serves as the
reference at normal gravity (Fig.~\ref{fig:config}a). Pure hydrogen issues from a round tube of inner diameter
$d=5$\,mm at a bulk velocity $U_0=7.7$\,m/s (9 standard liters per minute), so that the Reynolds number based on the
kinematic viscosity of hydrogen is $\mathrm{Re}=U_0d/\nu_\mathrm{F}=347$. The tube is surrounded by an air co-flow of
160\,mm diameter at $\Uco=0.15$\,m/s, and both streams enter at 300\,K and atmospheric pressure. At normal gravity,
$g_0=9.81$\,m/s$^2$, the jet velocity greatly exceeds the buoyant velocity scale $(g_0d)^{1/2}=0.22$\,m/s, whereas the
co-flow velocity is comparable to it: the jet Froude number $U_0^2/(g_0d)$ is $1.2\times10^3$, and the co-flow Froude
number $\Frco=\Uco^2/(gd)$ is 0.46.

Dennis et al.\ measured the temperature by chirped-probe-pulse fs-CARS at 5\,kHz at 42 points between $x/d=1$ and 16,
$x$ being the axial distance from the burner exit, and reported radial profiles of the mean and root-mean-square (rms)
temperature together with temperature time series~\cite{Dennis2016}. These data were digitized from the publication; the measured flicker frequency is taken from the peak of the Fourier
spectrum of the time series (supplementary material).

\subsection{Flow equations and thermochemical closures}
\label{sec:closures}
The flow is described by the variable-density Navier--Stokes equations in the low-Mach-number limit, in cylindrical
coordinates $(x,r,\theta)$ with gravity acting in the $-x$ direction, so that buoyancy enters through the density
field. Combustion is represented by a steady flamelet model~\cite{Peters1984}: a transport equation is solved for the
mixture fraction $\xi$, and the density, temperature, viscosity, and diffusivity are read from flamelet tables. When
radiative heat loss is included, a second scalar carries the enthalpy defect (Section~\ref{sec:radiation}).

Two thermochemical closures are used. The unity-Lewis-number closure (ULe) is used for all radiating simulations and
for the series that locate the limit of the flicker, because the present enthalpy-defect formulation assumes equal diffusivities of the mixture fraction and the
enthalpy. At normal gravity, the differential-diffusion
closure DD-eff and ULe are compared over a range of co-flow velocities, and the adiabatic flames at 0.05 and
$0.1\,g_0$ are repeated with DD-eff to show how differential diffusion changes the onset and the limit of the
flicker.

The tables are generated from steady counterflow diffusion flames of hydrogen against air at a strain rate of
$a=50$\,s$^{-1}$, with both streams at 300\,K and 0.993\,atm, computed with OPPDIF~\cite{Lutz1997} and the GRI-Mech 3.0
mechanism without its nitrogen chemistry~\cite{GRIMech30}. For ULe, all species diffuse with the thermal diffusivity
$\alpha$, and the table is parameterized by the mixture fraction, which is transported with $\alpha$. For DD-eff, the
flamelets are computed with mixture-averaged diffusion without thermal diffusion (the Soret effect). Mixture fractions defined from
different elements then no longer coincide, and the table is parameterized by Bilger's mixture fraction
$\xi_\mathrm{B}$~\cite{Bilger1988}, which is transported with its effective diffusivity $D_\mathrm{eff}(\xi_\mathrm{B})$
obtained from the flamelet solutions, 1.77 times $\alpha$ at stoichiometry ($\xi_\mathrm{st}=0.0284$). Two further
closures and the effect of the strain rate of the tables are examined in the supplementary material.

\subsection{Radiative heat loss}
\label{sec:radiation}
Radiation removes enthalpy from the hot gas, which lowers its temperature, raises its density, and so reduces its
buoyancy. The loss is represented by the enthalpy defect $\Delta h=h-h_\mathrm{m}(\xi)$, where $h_\mathrm{m}(\xi)$ is
the enthalpy of adiabatic mixing of the two streams, linear in $\xi$. With unity Lewis numbers, mixture fraction and
enthalpy diffuse at the same rate, and the defect obeys
\begin{equation}
\frac{\partial(\rho\Delta h)}{\partial t}+\nabla\cdot(\rho\mathbf{u}\Delta h)=\nabla\cdot(\rho\alpha\nabla\Delta h)
-\rho q_\mathrm{rad},\qquad \rho q_\mathrm{rad}=4\sigma\kappa_\mathrm{P}\left(T^4-T_\infty^4\right),
\label{eq:defect}
\end{equation}
where the optically thin radiative loss of water vapor is evaluated with its Planck-mean absorption coefficient
$\kappa_\mathrm{P}$, in m$^{-1}$ at the local partial pressure, from the TNF radiation model~\cite{TNFrad,Barlow2001}, $\sigma$ is the Stefan--Boltzmann constant, and
$T_\infty=300$\,K. The flamelet table is extended by a normalized defect $\eta=-\Delta h/\Delta H_\mathrm{max}(\xi)\in
[0,1]$, where $\Delta H_\mathrm{max}(\xi)$ is the enthalpy difference between the adiabatic flamelet and the same
composition at 300\,K. Along this coordinate, the composition is kept at its adiabatic flamelet value: the temperature
follows from the enthalpy at fixed composition, and the density and transport properties are evaluated at this
temperature with Cantera~\cite{Cantera}. Comparisons with radiative counterflow flames and the energy balance of a
steady reduced-gravity flame are given in the supplementary material. The radiative cooling time of the products, the enthalpy they can lose divided by their radiative loss rate, is
0.26--0.35\,s.

Two modifications of the sink separate the effects of radiation. In the first, the emission is multiplied by a
constant factor $\chi\ge0$: $\chi=1$ is the model above, $\chi=0$ the adiabatic flame, and factors above 1 are model
experiments with a stronger emission. In the second, the radiative source per unit mass, $q_\mathrm{rad}$, is frozen in
space at its distribution in a given state of the flame. The steady loss is then unchanged, but the source no longer
responds when a perturbation displaces gas of different temperature and composition, so that a comparison of frozen
and responding sources isolates this response.

The optically thin model neglects the reabsorption of the emitted radiation. Its effect is estimated by
post-processing the steady radiating flame at $0.1\,g_0$: the radiative transfer equation is solved by ray tracing
with the rank-correlated spectral-line weighted-sum-of-gray-gases model~\cite{Solovjov2017} and the absorption-line
blackbody distribution function of water vapor from the HITEMP 2010 database~\cite{Pearson2014,Rothman2010}. The net
radiative loss of the gas above 1000\,K is 0.80 times the optically thin sink of the simulations near the burner and
0.72 times at $x/d=80$, so that the net loss of the flame corresponds to $\chi\approx0.72$--0.80 (supplementary
material).

\subsection{Numerical procedure and simulation cases}
\label{sec:numerics}\label{sec:cases}
The equations are solved with a finite-difference code originally developed by Pierce and Moin~\cite{PierceMoin2004} and extended for the present study. It uses second-order central differences for the momentum equations,
the QUICK scheme for the scalars, and a second-order semi-implicit time integration at an axial Courant number of 1.0
(supplementary material). At $\mathrm{Re}=347$, the flow is laminar and
resolved by the grid: the dynamic Smagorinsky model retained in the code yields an eddy viscosity below 1.3\% of the
molecular viscosity in 99\% of the cells.

At normal gravity, the domain extends over $30d$ in the axial and $10d$ in the radial direction, with $144\times108$
cells in $x$ and $r$ (minimum spacing $0.034d$ and $0.014d$) and four azimuthal planes. Both streams enter with
top-hat velocity profiles, and convective conditions are applied at the outlet and at the lateral boundary. At reduced
gravity, the flames become longer and wider, and their oscillation can recede far downstream. These flames are computed
on a domain of $94d\times30d$ ($240\times162$ cells), with the co-flow confined to $r\le16d$ as in the experiment and a
quiescent inlet beyond (Fig.~\ref{fig:config}b). The grid and the domain were checked by doubling the resolution, by
using 8 or 16 azimuthal planes, and by enlarging the domain to $62d$ in length or $30d$ in width at normal gravity and to
$158d$ or $45d$ at reduced gravity. These changes alter the frequency of a sustained oscillation by at most 3\% and the mean flame position by at most $0.03d$. Apart from two domain effects discussed in Sections~\ref{sec:lowg}
and~\ref{sec:coflow}, they do not change whether an oscillation is sustained or recedes. The speed at which an oscillation recedes changes by at most 5\% with the domain. Doubling the radial resolution
matters more: the flame with $\chi=0.75$ at $0.1\,g_0$, continued on the finer grid, recedes 26\% faster. On 16 planes, the
reduced-gravity flames stay axisymmetric, even when seeded with inflow noise (supplementary material).

The simulations are listed in Table~\ref{tab:cases}. At normal gravity, each simulation starts from the limit cycle of
the reference flame, and statistics are collected over 13--29 flicker periods once the flame and the co-flow are
statistically stationary; no perturbation is imposed at the inflow. The reduced-gravity simulations start from the
limit cycle of the ULe flame at normal gravity on the $94d\times30d$ domain, and gravity is reduced in one step at
$t=0$, as at the start of a drop-tower experiment. Steps to gravity levels between 0.01 and $0.378\,g_0$, with and
without radiation, show at which gravity the flicker is lost.

Two series isolate the radiative loss and its response. At $0.1\,g_0$, the emission is switched on in the oscillating
adiabatic flame, with factors $\chi$ between 0.5 and 1, and switched off in the radiating flame that has lost its
flicker. A short gravity pulse applied to the steady radiating flame excites a wave packet, whose evolution is measured
relative to the same flame without the pulse. Repeating this comparison with the radiative source frozen isolates the
contribution of the source response.

Changing the co-flow at normal and reduced gravity separates its effect from that of gravity. The flames with
$\chi=1.15$ and 1.65 at $0.165\,g_0$ were computed after the limit relation of Section~\ref{sec:limit} had been calibrated, and they serve as its test. With DD-eff, the flame at $0.1\,g_0$ starts from the DD-eff flame at normal gravity on the
same domain, and the flame at $0.05\,g_0$ from that at $0.1\,g_0$ (4.04\,s after its step).

\begin{table*}[!tbp]
\centering\footnotesize
\caption{Simulation cases. Domains: $30d\times10d$ at normal gravity and $94d\times30d$ at reduced gravity, unless
noted.}
\label{tab:cases}
\begin{tabular}{p{0.18\textwidth}p{0.29\textwidth}p{0.17\textwidth}p{0.12\textwidth}p{0.14\textwidth}}
\toprule
\RR Case series & \RR $g/g_0$ & \RR $\Uco$ (m/s) & \RR Closure & \RR Radiation \\
\midrule
\RR Reference flame & \RR 1 & \RR 0.15 & \RR ULe & \RR no, yes \\
\RR  & \RR 1 & \RR 0.15 & \RR DD-eff & \RR no \\
\RR Co-flow at $g_0$ & \RR 1 & \RR 0.15, 0.18, 0.35, 0.5, 0.6, 0.7, 0.8$^\mathrm{a)}$ & \RR ULe & \RR no \\
\RR  & \RR 1 & \RR 0.12, 0.15, 0.18, 0.21 & \RR DD-eff & \RR no \\
\RR Gravity step & \RR 0.01, 0.02, 0.03, 0.05, 0.1, 0.165, 0.25, 0.378 & \RR 0.15 & \RR ULe & \RR yes \\
\RR  & \RR 0.02, 0.05, 0.1, 0.165, 0.25 & \RR 0.15 & \RR ULe & \RR no \\
\RR  & \RR 0.05, 0.1 & \RR 0.15 & \RR DD-eff & \RR no \\
\RR Co-flow at low $g$ & \RR 0.25 & \RR 0.25 & \RR ULe & \RR yes, no \\
\RR  & \RR 0.25 & \RR 0.35 & \RR ULe & \RR no \\
\RR  & \RR 0.05, 0.1 & \RR 0.075 & \RR ULe & \RR yes \\
\RR Emission factor & \RR 0.1 & \RR 0.15 & \RR ULe & \RR $\chi=0$--1 \\
\RR Gravity pulse$^\mathrm{b)}$ & \RR 0.05, 0.1 & \RR 0.15 & \RR ULe & \RR responding, frozen \\
\RR Test of the limit & \RR 0.165 & \RR 0.15 & \RR ULe & \RR $\chi=1.15$, 1.65 \\
\bottomrule
\end{tabular}
\par\smallskip{\footnotesize $^\mathrm{a)}$0.5--0.7\,m/s on $62d\times10d$ and 0.7, 0.8\,m/s on $94d\times30d$.
$^\mathrm{b)}$$+0.05$ or $+0.025\,g_0$ for 33.6\,ms at $0.05\,g_0$; $+0.025\,g_0$ at $0.1\,g_0$.}
\end{table*}

\subsection{Characterization of the oscillation}
\label{sec:diagnostics}
The oscillation is characterized by the outer flame edge $R(x,t)$, the largest radius of the 1000\,K isotherm in the
diametral plane. Its local swing is obtained by subtracting from $R$ its running mean over one period and taking the peak-to-peak value of
the remainder in windows of 1.25 periods. The onset of the oscillation, $x_\mathrm{on}$, is the most upstream position at
which the swing exceeds $0.25d$ in the last 0.5\,s of a simulation; during a transient, the most upstream position at
which this threshold is exceeded at a given time is called the trailing front. The flicker frequency is obtained at
normal gravity from the zero crossings of the temperature time series at the measurement points, and at reduced
gravity from the autocorrelation of $R$. Times are given in s or in units of $d/U_0=0.65$\,ms.

The state of a simulation is decided from the swing in successive 0.5\,s windows within $x\le L-10d$, $L$ being the
length of the domain; where the co-flow has been changed, only windows after it has been re-established are used
(supplementary material). The criteria below are applied in the order lost, receding, and sustained; the first one met
determines the state. The oscillation is lost when
its swing stays below $0.1d$ everywhere in the last window, or falls by more than 30\% per window over the last 1\,s.
It is receding when its onset moves downstream by more than $2d$ over the last 2\,s. It is sustained when its largest
swing exceeds $0.25d$ in each of the last two windows and changes by less than 15\% between them, its onset moving by
less than $2d$. A simulation that meets none of these criteria is undecided.

Radiation changes the temperature field directly, so the oscillation at reduced gravity is also followed in quantities
that radiation changes only through the flow: the outer radius of the stoichiometric surface $\xi=\xi_\mathrm{st}$, the
peak azimuthal vorticity outside the temperature maximum, in the outer shear layer, and the radial velocity at $r=5d$.
The hot outer layer (hereafter the hot layer) is the gas heated by the flame and accelerated by buoyancy outside the
temperature peak. It is characterized by $\Delta U$, the difference between the time-averaged axial velocities at the
radius of peak temperature and in the co-flow at $r=12d$, and by the ratio $\Delta U/\Uco$, both defined where the
temperature peak lies off the axis. The time-averaged profiles of this layer are the base states of the stability
analysis.

\subsection{Local linear stability analysis}
\label{sec:lsa}
Local stability analysis relates the profiles of the hot layer to the growth of disturbances and to their upstream
spreading. Under the parallel-flow approximation, the base state at each axial station consists of the radial profiles
of the axial velocity $U(r)$, the density $\rho(r)$, and the viscosity $\mu(r)$: the time-averaged profiles for
oscillating flames, and the profiles of the steady flame for flames without flicker. Two models are used. A simplified model describes the hydrodynamics of the hot layer; it gives its absolute or convective character and the speed at which disturbances spread upstream. A complete linearization of the flamelet model adds the dilatation, the coupling of the density to the composition, and the radiative source
with its response to the perturbation. The radiative response is switched on and off on the same fixed flows and mode
branch, and the difference measures how it changes the growth rates and the spreading speed. Only this difference is used: the terms it adds change the absolute instability of the layer more than the radiative
response does, so the complete linearization does not decide whether a layer is absolutely unstable (supplementary
material). Whether an oscillation is sustained is decided by the simulations.

In the simplified model, perturbations of the form $\hat{q}(r)\exp[\mathrm{i}(kx+m\theta-\omega t)]$ obey the linearized variable-density equations with a
solenoidal perturbation velocity
\begin{align}
\rho\,(-\mathrm{i}\omega+\mathrm{i}kU)\,\hat{\mathbf{u}}+\rho\,\frac{\mathrm{d}U}{\mathrm{d}r}\,\hat{v}\,\mathbf{e}_x
+\hat{\nabla}\hat{p}&=\mu\,\hat{\nabla}^2\hat{\mathbf{u}}-g\,\hat{\rho}\,\mathbf{e}_x, \label{eq:lsa_mom}\\
\hat{\nabla}\cdot\hat{\mathbf{u}}&=0, \label{eq:lsa_cont}\\
(-\mathrm{i}\omega+\mathrm{i}kU)\,\hat{\rho}+\frac{\mathrm{d}\rho}{\mathrm{d}r}\,\hat{v}&=\kappa\,\hat{\nabla}^2\hat{\rho},
\label{eq:lsa_rho}
\end{align}
where $\hat{\mathbf{u}}=(\hat{u},\hat{v},\hat{w})$, the operators $\hat{\nabla}$ and $\hat{\nabla}^2$ follow from
$\partial/\partial x\to\mathrm{i}k$ and $\partial/\partial\theta\to\mathrm{i}m$, and $\kappa=\mu/(\rho\,\mathrm{Pr})$
with $\mathrm{Pr}=0.7$. The model retains the inertial and buoyancy terms of the hot layer but neglects the dilatation
produced by molecular transport and heat release in the perturbation, the coupling of the density to the composition,
the radiative source, and the radial variation of the viscosity in the viscous terms.

The absolute frequency $\omega_0=\omega(k_0)$, or $f_0=\mathrm{Re}\,\omega_0/2\pi$ in Hz, is evaluated at the saddle point $\partial\omega/\partial k(k_0)=0$ of the
branch of the hot layer, and the flow is absolutely unstable where $\omega_{0,i}>0$~\cite{HuerreMonkewitz1990}. In a
convectively unstable flow, the trailing edge of the impulse response travels at the velocity $V_-$ at which the growth
rate along the ray $x/t=V$ vanishes~\cite{HuerreMonkewitz1990}. This velocity is measured in the laboratory frame; the
upstream spreading speed relative to the co-flow is $S=\Uco-V_-$, so that the layer is absolutely unstable where
$S>\Uco$.

Equations~\eqref{eq:lsa_mom}--\eqref{eq:lsa_rho} are discretized by Chebyshev collocation on $0\le r\le30d$ with
$N=80$ points, half of them below $r=4d$, regularity conditions on the axis~\cite{Khorrami1989}, and vanishing
perturbations at $r=30d$. The mode of the hot layer is identified by its phase velocity and by the radial position of
its eigenfunction maximum, and every saddle point used was verified to be a pinch point by the Briggs--Bers
criterion~\cite{HuerreMonkewitz1990} (supplementary material). At reduced gravity, the results depend on the outer
radius; they were checked with outer radii of $40d$ and $50d$, and with $N=120$. For the adiabatic flames at 0.05 and
$0.1\,g_0$, whose results change between $30d$ and $40d$, the values at $40d$ are used.

%=====================================================================================================================
\section{Results and discussion}
\label{sec:results}

\subsection{Reference flame and normal-gravity flicker}
\label{sec:flame1g}
\begin{figure*}[!tbp]
\centering
\figfile{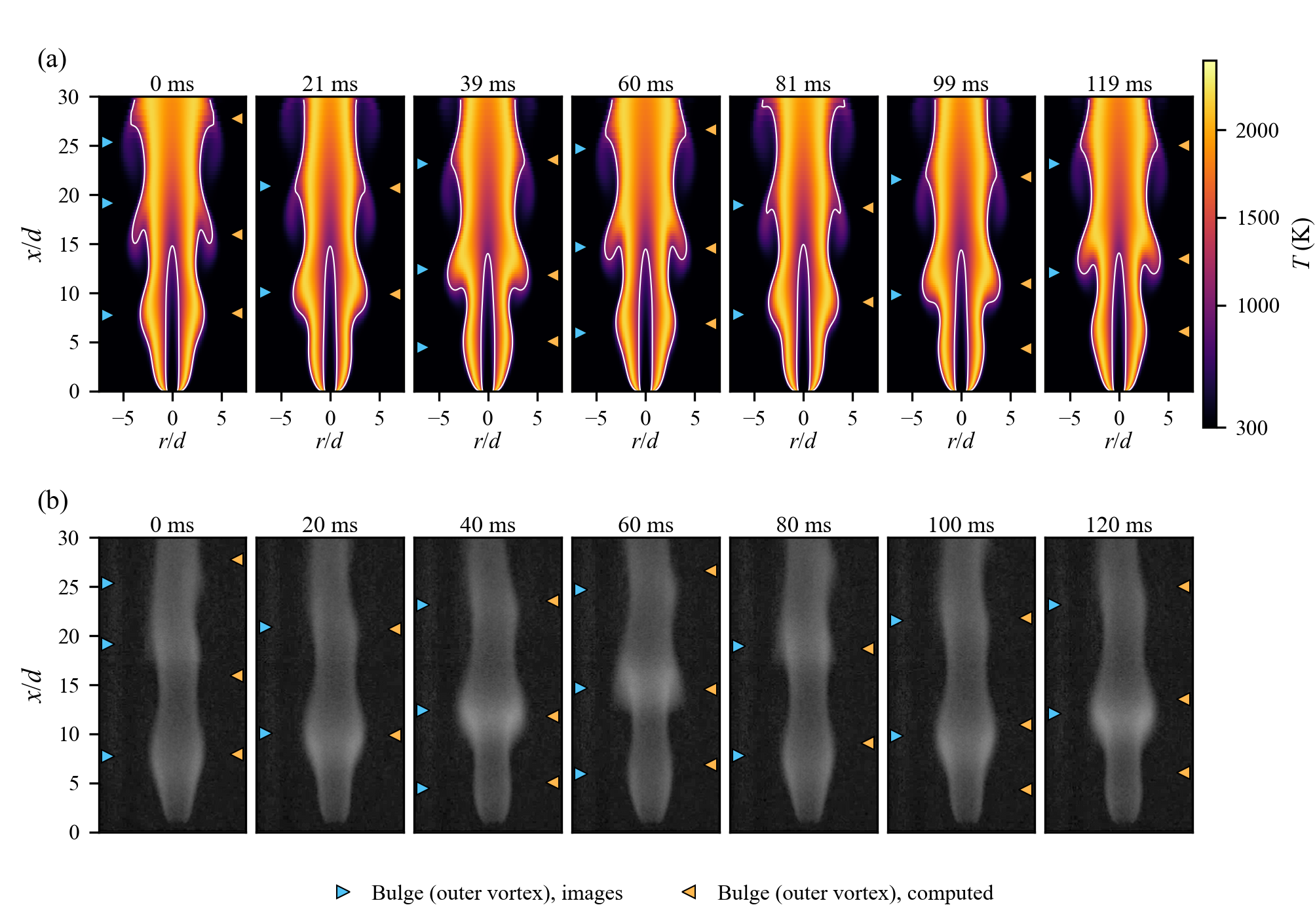}
\caption{Flame at normal gravity over almost two flicker periods: (a) temperature in the diametral plane computed
with ULe (white: 1000\,K isotherm); (b) high-speed images of Ref.~\cite{Dennis2016} (a U.S. Government work in the
public domain), scaled and aligned in phase by a fit to the luminous width of the flame. Triangles: bulge heights from the images (cyan) and
the computation (orange).}
\label{fig:sequence}
\end{figure*}

At normal gravity, the computed flame reproduces the flicker seen in the high-speed images of the
experiment~\cite{Dennis2016} (Fig.~\ref{fig:sequence}). In both, the flame is attached to the burner, and axisymmetric
bulges of its outer edge form near $x/d\approx5$ about every 70\,ms, travel downstream by about $3d$ every 20\,ms while they grow, and neck
the flame between them; the computed and observed bulge positions differ by $1.2d$
(rms) over the 120\,ms shown. No perturbation is imposed at the inflow, yet both closures settle on a periodic limit
cycle, whose peak-to-peak temperature fluctuation varies by less than 1\% between successive four-period windows: the
flicker is a self-sustained oscillation of the flame. Its edge oscillates as a single traveling wave, whose amplitude
grows from about $0.01d$ at $x/d=0.5$ to 0.9--1.3$d$ at $x/d=12$--16 (supplementary material).

The two closures reproduce different features of the measured flame (Fig.~\ref{fig:validation}; values in
the supplementary material). ULe matches the measured frequency of 14.03\,Hz to within 3.4\% and gives the peak
temperature closest to the measurements, but places the flame 0.22--0.44$d$ inside the measured one at $x/d=1$--5.
DD-eff widens the flame by 0.30--0.49$d$ near the burner and raises the peak mean temperature by about 240\,K. It
matches the measured flame position near the burner to within 0.05--0.21$d$ but is too wide further downstream, with
an outer 1000\,K radius of $3.52d$ against $2.65d$ measured at $x/d=8$. Its flicker frequency is 8.4\% below the measured value. Both closures overpredict the swing of the flame edge at $x/d=4$--8, by a factor of 1.7--3.1 relative to
the swing inferred from the measured temperature fluctuations (Fig.~\ref{fig:validation}e; supplementary material).
Across the closures, the frequency decreases with the flame size, close to the inverse-square-root dependence
expected for buoyancy-driven flicker~\cite{CetegenAhmed1993}, and a closure that places the flame too far inward also predicts too
high a frequency (supplementary material): agreement in frequency alone does not identify the thermochemical closure.
At six selected measurement points with a periodic signal, the ULe simulation, shifted in time, follows the period and the waveform of the measured oscillation, with correlation
coefficients of 0.49--0.79 (supplementary material).

\begin{figure*}[!tbp]
\centering
\figfile{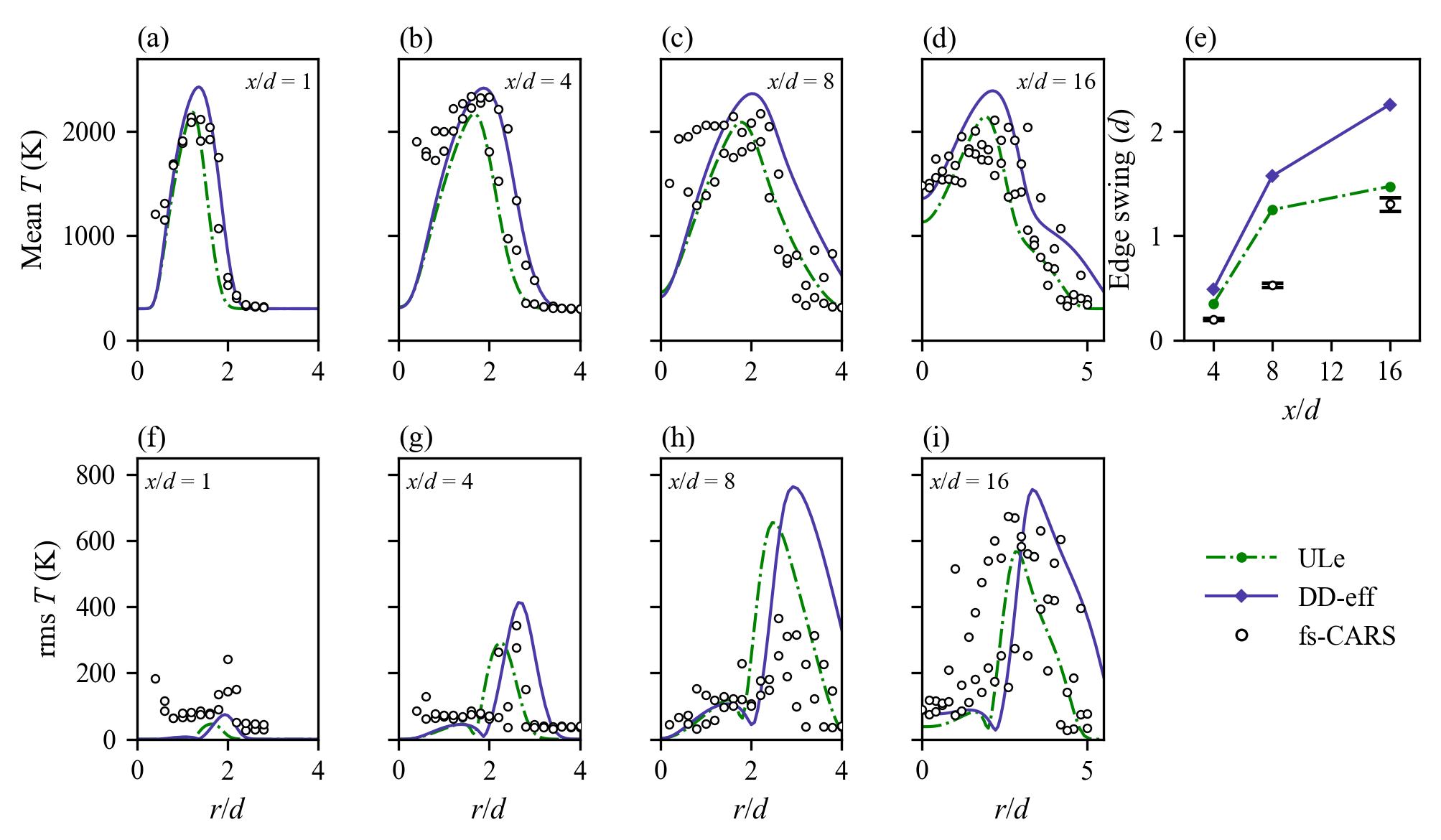}
\caption{Comparison with the fs-CARS measurements~\cite{Dennis2016} at normal gravity for ULe and DD-eff: (a--d) mean
and (f--i) rms temperature at $x/d=1$, 4, 8, and 16; (e) swing of the outer 1000\,K edge against $x/d$, with the range
inferred from the measurements. Symbols: measurements, digitized from the published profiles; the measured rms
includes single-shot noise.}
\label{fig:validation}
\end{figure*}

The flicker at normal gravity is sensitive to the co-flow but hardly to radiation. From 0.15 to 0.18\,m/s, the
frequency rises by 4.0\% with ULe and by 4.1\% with DD-eff, more than the deviation of ULe from the measurement, so
that flicker frequencies can be compared with measurements only when the co-flow velocity is accurately known.
Radiation removes 5.3\% of the fuel power, lowers the peak temperature near the burner by 15--43\,K, and raises the
frequency by only 1.2\% (supplementary material).

\subsection{Radiative suppression and recovery of flicker at reduced gravity}
\label{sec:lowg}\label{sec:radeffect}
\begin{figure*}[!tbp]
\centering
\figfile{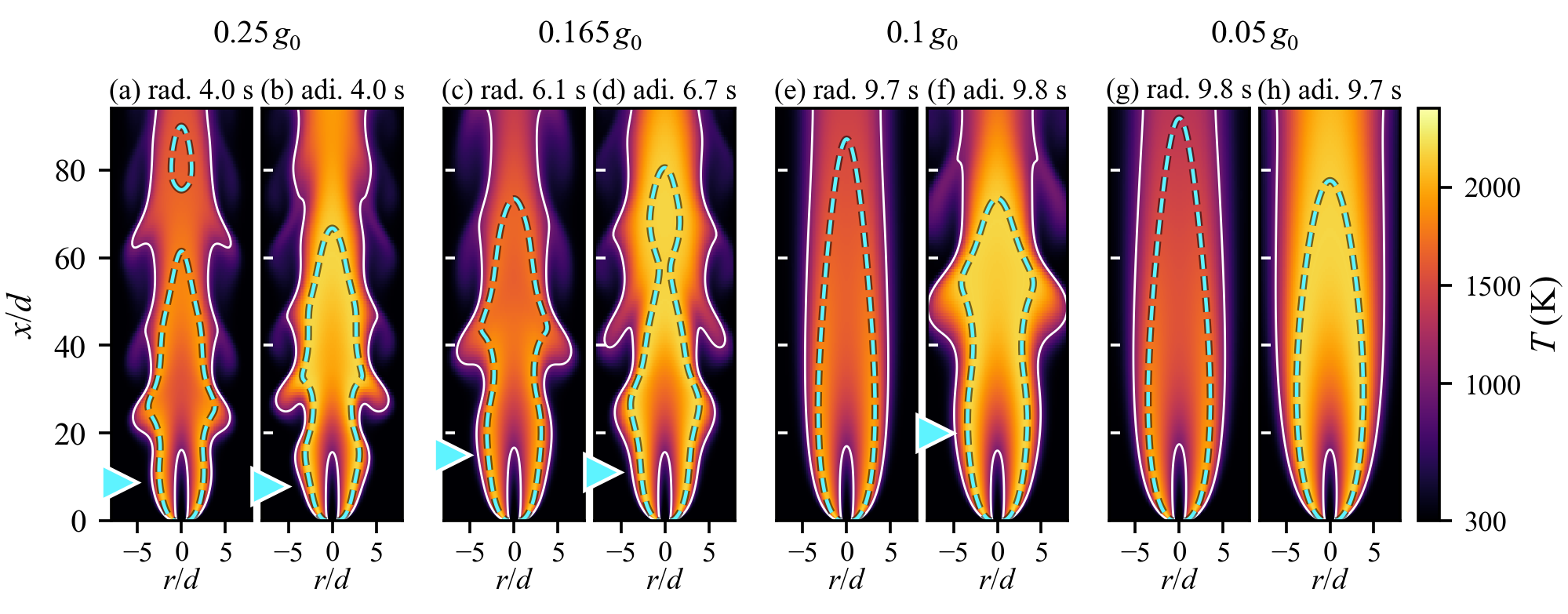}
\caption{Temperature in the diametral plane at the end of each record after the gravity step (time above each
panel), radiating (rad.) and adiabatic (adi.) flames at (a, b) 0.25, (c, d) 0.165, (e, f) 0.1, and (g, h)
$0.05\,g_0$; axial scale compressed twofold. White: 1000\,K isotherm; cyan dashed: stoichiometric surface; triangles:
onset of the sustained oscillation.}
\label{fig:lowg}
\end{figure*}

\begin{figure*}[!tbp]
\centering
\figfile{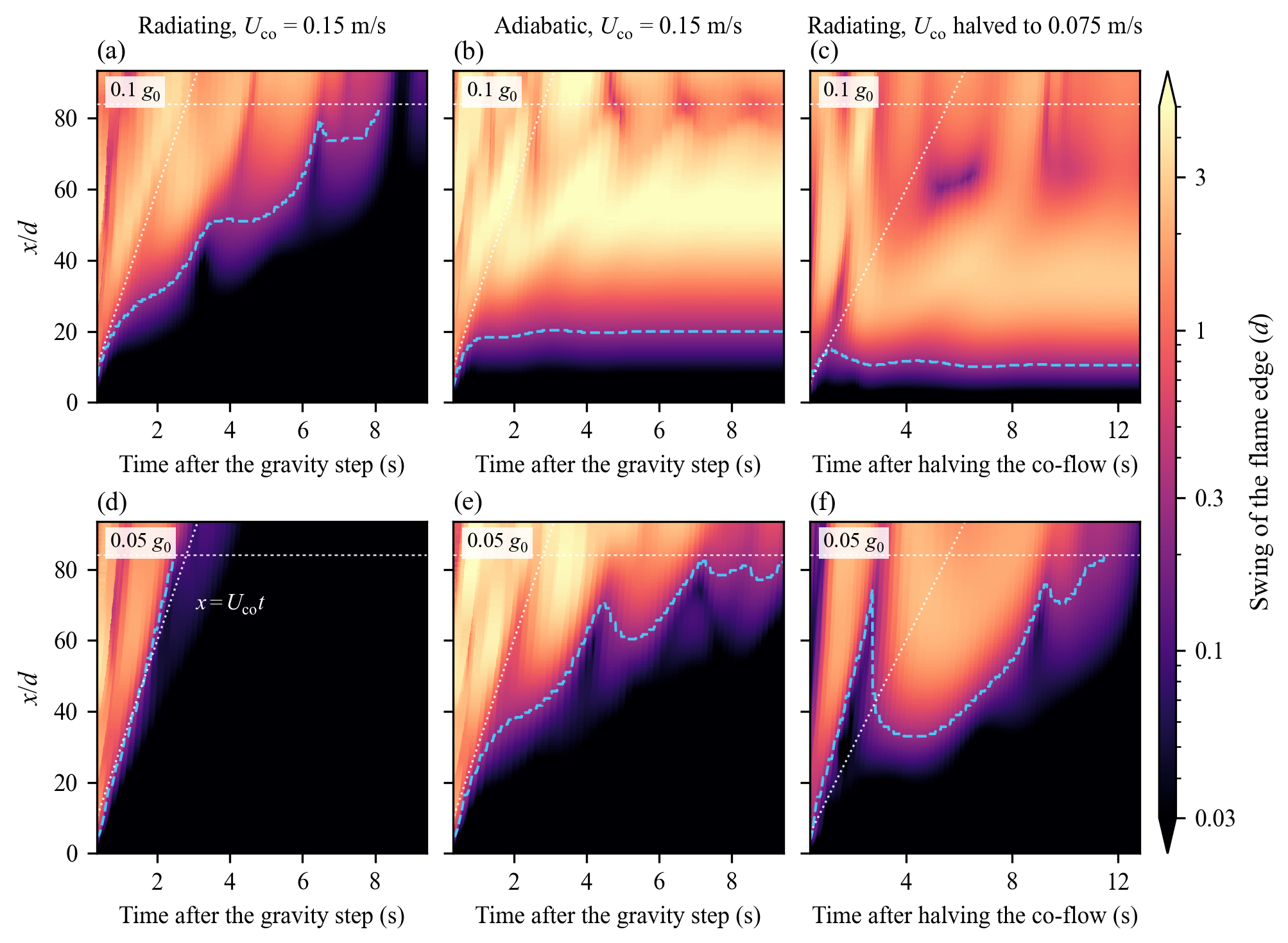}
\caption{Swing of the flame edge after a step reduction of gravity to $0.1\,g_0$ (a--c) and $0.05\,g_0$ (d--f):
(a, d) radiating and (b, e) adiabatic flames at $\Uco=0.15$\,m/s; (c, f) radiating flames after the co-flow is halved
to 0.075\,m/s, 4.4 and 3.2\,s after the step. Cyan dashed: trailing front (swing $0.25d$); white dotted: $x=\Uco t$;
white dashed: $x/d=84$.}
\label{fig:spacetime}
\end{figure*}

Reducing gravity changes the role of radiation. After a step reduction of gravity, the radiating flame loses its flicker at $0.1\,g_0$, the adiabatic flame not until $0.05\,g_0$ (Table~\ref{tab:lowg}). As gravity decreases,
the flame widens and lengthens (Fig.~\ref{fig:lowg}): at $0.05\,g_0$, the outer 1000\,K radius at $x/d=30$ grows from
$2.6d$ to about $5d$, and the stoichiometric contour reaches the axis at $x/d=84$--94. The oscillation responds to the
step in two stages (Fig.~\ref{fig:spacetime}a, b, d, e). The normal-gravity flicker near the burner decays first, and
its last bulges, together with the disturbance created by the step, form a wave packet of longer wavelength and lower
frequency that is carried downstream. Whether an oscillation then remains depends on gravity and radiation. At
0.165--0.378\,$g_0$ with radiation, and down to $0.1\,g_0$ without it, the trailing front of the packet stops and the
oscillation settles on a limit cycle, its onset steady to within $1d$. At lower gravity, and at $0.1\,g_0$ with
radiation, the front keeps moving downstream, at 0.4--2.1 times the co-flow velocity, and leaves a steady flame behind
it.

\begin{table*}[!tbp]
\centering\small
\caption{Flicker after changes of gravity and co-flow (ULe, $\Uco=0.15$\,m/s unless noted): state after the transient, onset
$x_\mathrm{on}$ in the final record, and frequency (in parentheses: of the transient packet that leaves the flame).
States by the rules of Section~\ref{sec:diagnostics}.}
\label{tab:lowg}
\begin{tabular}{cccccc}
\toprule
$g/g_0$ & Radiation & $\Frco$ & State & $x_\mathrm{on}$ ($d$) & $f$ (Hz) \\
\midrule
0.378 & yes & 1.21 & sustained & 6.4 & 8.56 \\
0.25 & yes & 1.83 & sustained & 8.8 & 6.70 \\
0.25 & no & 1.83 & sustained & 7.9 & 6.69 \\
0.165 & yes & 2.78 & sustained & 15.1 & 5.50 \\
0.165 & no & 2.78 & sustained & 11.1 & 5.31 \\
0.1 & yes & 4.59 & lost & -- & (4.53) \\
0.1 & no & 4.59 & sustained & 20.0 & 3.95 \\
0.1, DD-eff & no & 4.59 & sustained & 10.5 & 3.21 \\
0.05, DD-eff & no & 9.17 & receding & 37.0 & 2.41 \\
0.05 & yes & 9.17 & lost & -- & --$^\mathrm{b)}$ \\
0.05 & no & 9.17 & lost ($158d$) & -- & (3.50) \\
0.02 & no & 22.9 & lost & -- & (1.45) \\
0.01--0.03 & yes & 15--46 & lost & -- & --$^\mathrm{b)}$ \\
0.25, $\Uco=0.25$ & yes & 5.10 & sustained & 29.7 & 8.56 \\
0.25, $\Uco=0.25$ & no & 5.10 & sustained & 15.6 & 7.00 \\
0.25, $\Uco=0.35$ & no & 10.0 & receding & 49.7 & 9.06 \\
0.1, $\Uco=0.075$ & yes & 1.15 & sustained & 10.5 & 3.67 \\
0.05, $\Uco=0.075$ & yes & 2.29 & undecided$^\mathrm{a)}$ & -- & (3.08) \\
\bottomrule
\end{tabular}
\par\smallskip{\footnotesize $^\mathrm{a)}$Receding beyond $x/d=84$ at the end of the record. $^\mathrm{b)}$Not
determined: the packet passes each station within fewer than three periods.}
\end{table*}

\begin{figure*}[!tbp]
\centering
\figfile{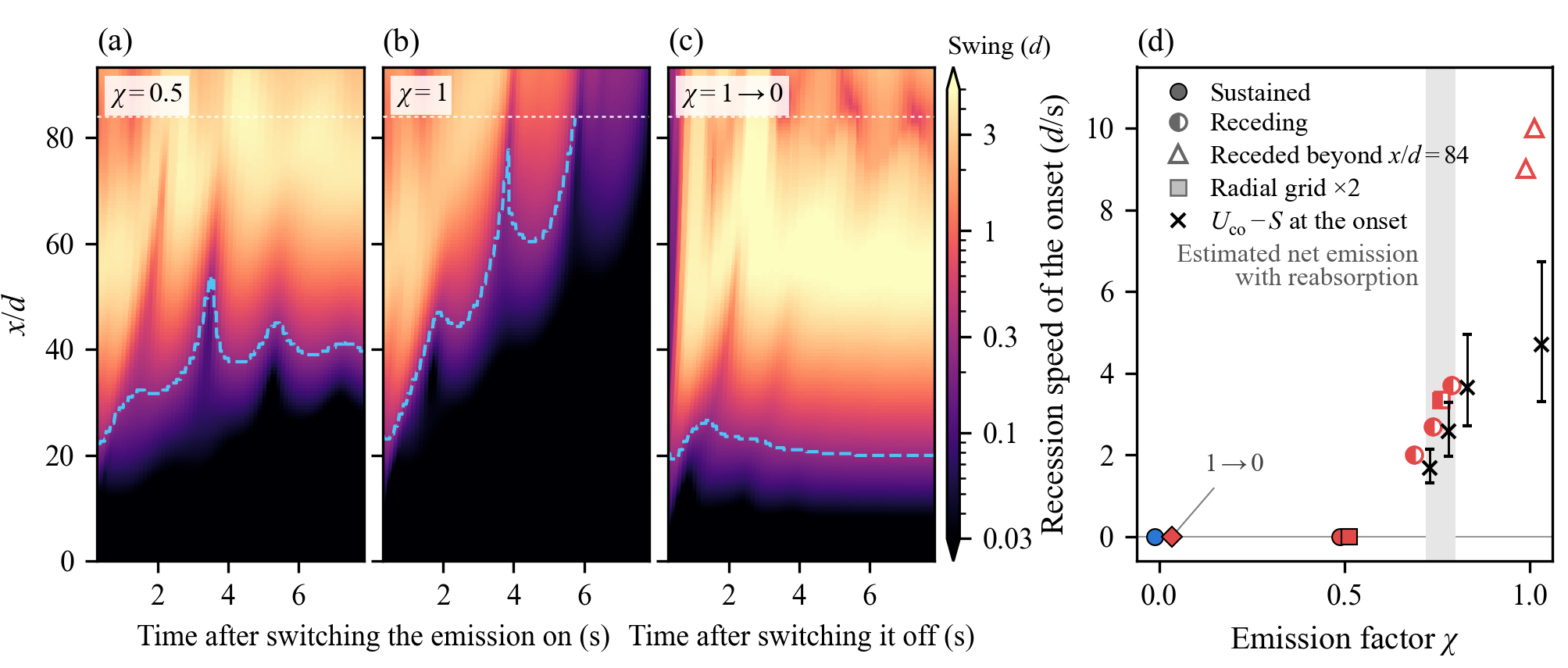}
\caption{Control of the flicker by the radiative emission at $0.1\,g_0$: swing of the flame edge after the emission
is switched on in the adiabatic flame with (a) $\chi=0.5$ and (b) $\chi=1$, and (c) after it is switched off in the
radiating flame (lines as in Fig.~\ref{fig:spacetime}); (d) recession speed of the onset against $\chi$ (crosses:
$\Uco-S$ from the local analysis at the onset, Section~\ref{sec:coflow}; bars: its range over the onset positions of
the last 2\,s; shaded: net radiative loss estimated with reabsorption).}
\label{fig:dose}
\end{figure*}

At fixed gravity and co-flow, the radiative loss alone decides whether the flame flickers (Fig.~\ref{fig:dose}). When
the emission is switched on in the oscillating adiabatic flame at $0.1\,g_0$, the oscillation recedes without
stopping, at about 9 diameters per second ($0.29\,\Uco$) after 3\,s, and after 5.9\,s, no swing above $0.25d$
remains within $x\le84d$. When the emission is switched off in the radiating flame that has lost its flicker, the
oscillation returns within a second, and over the last 2\,s of the 8.1\,s record, its onset stays at $x/d=20.0$, as in
the adiabatic flame.

The strength of the emission sets how fast the oscillation recedes (Fig.~\ref{fig:dose}d). With half the emission,
$\chi=0.5$, the onset moves from $x/d\approx20$ to about 38 during the first 3.5\,s and then holds its position, at
$x/d=39.0$--40.3 over the last 2\,s of the 8.1\,s record. With $\chi=0.7$, 0.75, and 0.8, it recedes at 2.0, 2.7, and 3.7
diameters per second, and with the full emission at about 9. The limit of the flicker at $0.1\,g_0$ thus lies between
$\chi=0.5$ and 0.7; on a grid with twice the radial resolution, the oscillation is likewise sustained at $\chi=0.5$ and
recedes at 0.75 and 1 (supplementary material). With
the emission scaled to the net loss estimated with reabsorption, 0.72--0.80 times the optically thin loss, the oscillation still recedes: the radiative loss of the flame lies just beyond the
limit of its flicker.

Quantities that radiation changes only through the flow show the same contrast: at $0.1\,g_0$,
from 3.5 to 6.0\,s after the step, the stoichiometric surface at $x/d=30$ swings by 0.48--0.50$d$ without radiation
and by 0.004--0.02$d$ with it, and at $0.05\,g_0$, the packet is 1.6--2.2 times weaker with radiation (supplementary
material). The adiabatic flame at $0.05\,g_0$ keeps a weak oscillation near the end of the $94d$ domain, but on a $158d$
domain this oscillation recedes and decays, so that the adiabatic flame, too, loses its flicker at this gravity
(supplementary material).

Where both flames keep the flicker, radiation displaces it downstream instead of removing it. At 0.15\,m/s, it moves the onset downstream, raising $x_\mathrm{on}$ by a factor of 1.11 at $0.25\,g_0$ and 1.36 at
$0.165\,g_0$, and it raises
the frequency by at most 4\%. At $0.25\,g_0$ with a co-flow of 0.25\,m/s ($\Frco=5.1$), the factor is 1.9, from
$x/d=15.6$ to 29.7. The displacement grows as gravity decreases and turns into suppression at $0.1\,g_0$. The frequency itself,
sustained or transient, follows the buoyant scaling $f\approx(12.5$--$14.3\,\mathrm{Hz})(g/g_0)^{1/2}$ between 0.1
and $0.378\,g_0$ (Table~\ref{tab:lowg})~\cite{CetegenAhmed1993,XiaZhang2018}.

Halving the co-flow restores the flicker of the radiating flame at $0.1\,g_0$ (Fig.~\ref{fig:spacetime}c), so that
radiation shifts a limit that gravity and co-flow set together. It can act on the flicker through the mean flow, whose
buoyancy it reduces, and through the disturbances, whose growth it damps.

\subsection{Radiative loss of buoyancy in the hot outer layer}
\label{sec:mechanism}
\begin{figure*}[!tbp]
\centering
\figfile{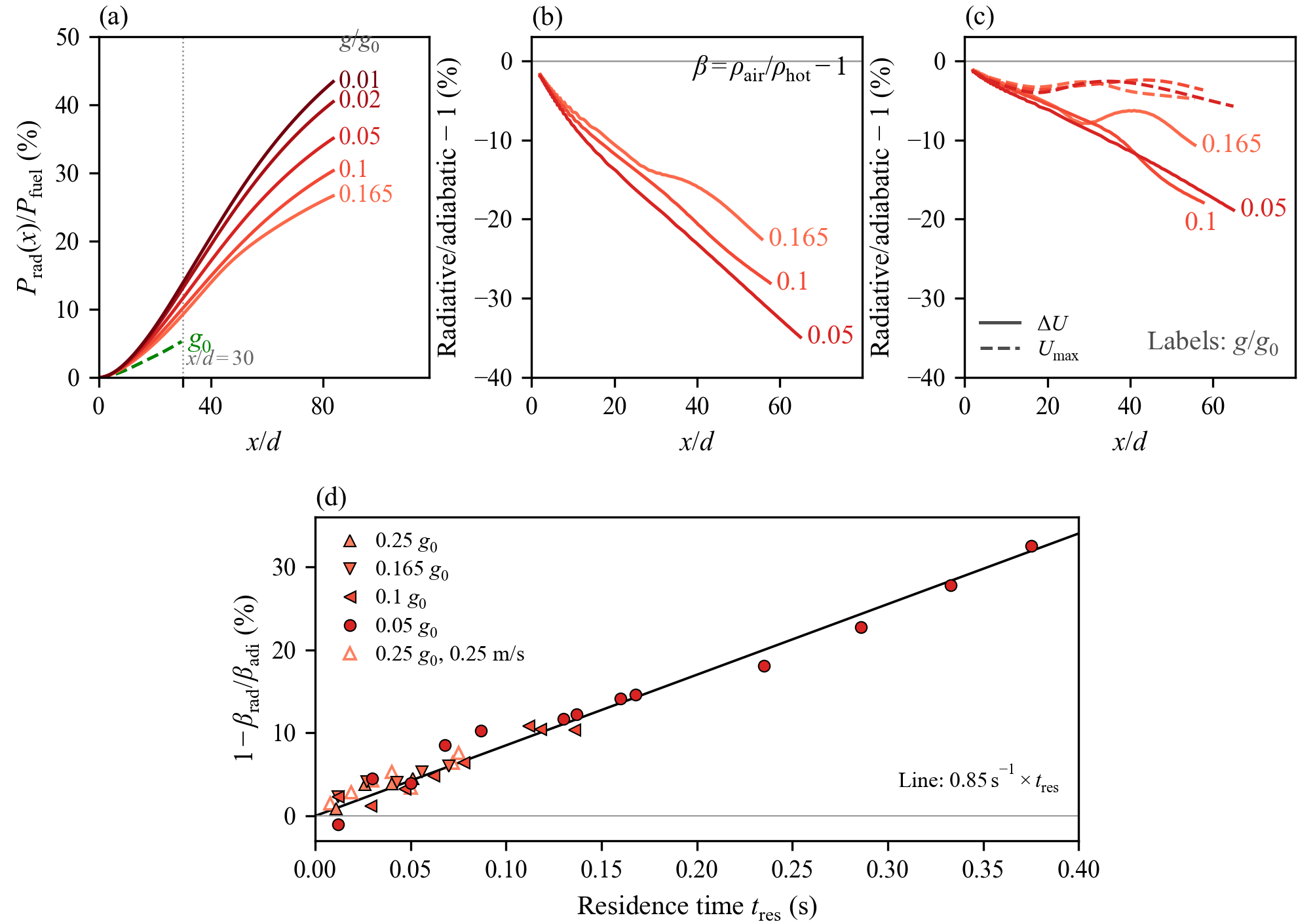}
\caption{Radiative loss and the hot layer at reduced gravity: (a) radiant power integrated from the burner to $x$, as
a fraction of the fuel power (labels: $g/g_0$; dashed: $g_0$, $30d$ domain); (b, c) radiating against adiabatic time-averaged flow:
(b) buoyancy factor $\beta=\rho_\mathrm{air}/\rho_\mathrm{hot}-1$, (c) velocity difference $\Delta U$ across the outer shear layer and largest axial velocity; (d) loss of buoyancy against the residence time (time means over the last 40\% of each record;
open: not used in the fit; line: Eq.~\eqref{eq:reslaw}).}
\label{fig:radflow}
\end{figure*}

Radiative cooling reduces the buoyancy of the hot outer layer. At reduced gravity, the radiative loss is no longer
small: at 0.1--0.01\,$g_0$, the flame radiates 10.3--14.1\% of the fuel power within $x/d\le30$, about twice as
much as at normal gravity, and 30--43\% up to $x/d=84$ (Fig.~\ref{fig:radflow}a). The gas stays longer in the flame:
it takes about 0.05\,s to pass the oscillating region at normal gravity ($x/d\le16$), but 0.4--0.5\,s to reach
$x/d=60$--90 at $0.05\,g_0$, longer than the radiative cooling time of the products, 0.26--0.35\,s.

The time-averaged flows over 4--6\,s after the step show how radiative cooling modifies the mean flow up to the end
of the hot layer, where the temperature peak reaches the axis (Fig.~\ref{fig:radflow}b, c). Radiation lowers the
peak temperature of the hot layer by up to 28\% at $x/d=30$--60. Because this gas is about one ninth as dense as the air, its density deficit $1-\rho_\mathrm{hot}/\rho_\mathrm{air}$ changes by only 2--5\%, but the buoyant
acceleration per unit mass, proportional to $\beta=\rho_\mathrm{air}/\rho_\mathrm{hot}-1$, falls by up to a third,
more where the gas has traveled longer. The
velocity difference across the outer shear layer decreases by 7--19\%, whereas the peak vorticity and the momentum
thickness of the layer change by at most 13\%. Radiation thus acts on the mean flow through the buoyancy that
accelerates the hot layer, and $\beta$, not the density deficit, measures this effect.

The loss of buoyancy follows the residence time of the gas in the layer (Fig.~\ref{fig:radflow}d). We take $\beta$ at the radius of peak temperature and define the residence time
$t_\mathrm{res}=\int_{0.5d}^{x}\mathrm{d}x'/U_\mathrm{hot}$, where $U_\mathrm{hot}$ is the axial velocity at that radius.
For each pair of radiating and adiabatic flames, we compare the annular hot layers upstream of the onset of the
adiabatic flame. The four pairs at 0.05--$0.25\,g_0$ and 0.15\,m/s then follow
\begin{equation}
1-\frac{\beta_\mathrm{rad}}{\beta_\mathrm{adi}}=\gamma\,t_\mathrm{res},\qquad \gamma=0.85\,\mathrm{s}^{-1},
\label{eq:reslaw}
\end{equation}
with an rms deviation of 1.25 percentage points for losses of up to 33\% at residence times of up to 0.38\,s. Fitted separately, the four pairs give rates of 0.84--0.99\,s$^{-1}$, and a fifth pair, at $0.25\,g_0$ and
0.25\,m/s, not used in the fit, follows Eq.~\eqref{eq:reslaw} with an rms deviation of 1.26 points. The rate $\gamma$ is
set by the emission of the flame products and changes little with gravity; the gravity dependence of the loss comes
from the residence time.

\subsection{Displaced-emission damping of flame disturbances}
\label{sec:response}
\begin{figure*}[!tbp]
\centering
\figfile{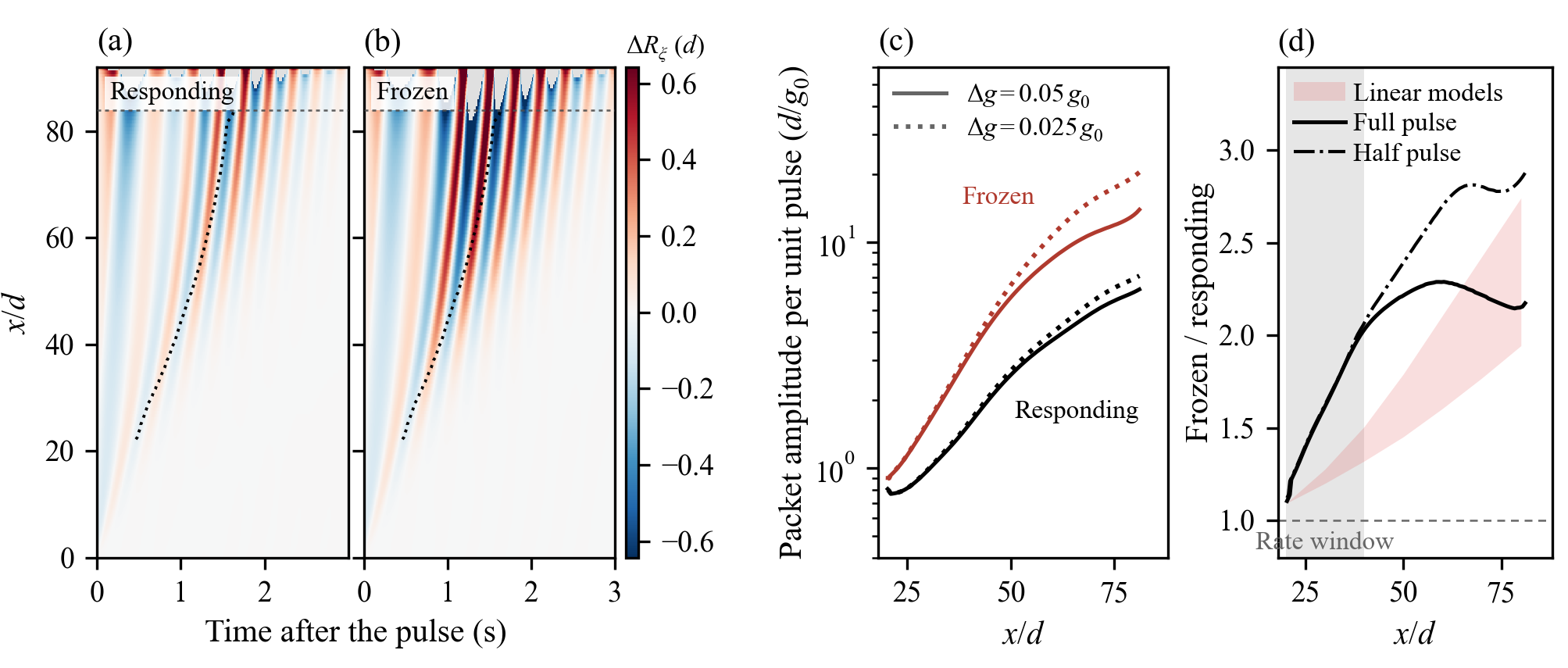}
\caption{Response of the steady radiating flame at $0.05\,g_0$ to a gravity pulse of 33.6\,ms: (a, b) displacement
of the stoichiometric surface from the unperturbed flame for a pulse of $0.05\,g_0$, with the radiative source (a)
responding and (b) frozen (dotted: packet time; gray: beyond the stoichiometric tip); (c) packet amplitude
per unit pulse; (d) ratio of the packet amplitudes with frozen and responding sources (gray: $x/d=20$--40) and the band of linear models that include the radiative
response, started from the simulated ratio at $x/d=20$. Curves in (c, d) end where the frozen packet breaks the stoichiometric surface.}
\label{fig:pulse}
\end{figure*}

Radiation also acts on the disturbances themselves. A flame that has lost its flicker is a linear amplifier of disturbances to its hot layer. A gravity pulse of $+0.05\,g_0$ lasting 33.6\,ms excites a wave packet in the steady radiating flame at
$0.05\,g_0$. The packet is convected downstream, and its largest displacement increases ninefold between $x/d=20$ and
84 (Fig.~\ref{fig:pulse}a). Behind the packet, the flame returns to its steady state within 3\,s,
and a pulse of half the strength produces the same packet per unit pulse, to within 3\% up to $x/d=40$
(Fig.~\ref{fig:pulse}c).

Freezing the radiative source separates its response from the mean-flow cooling. With the source per unit mass frozen at the steady state, the steady loss is unchanged, but the loss per unit mass is
held fixed at each location and no longer responds to the perturbations of temperature and composition that the wave
produces
(Fig.~\ref{fig:pulse}b). The initial displacement produced by the pulse is the same, but freezing the source raises the growth rate of the
packet along its path by 0.7--0.9\,s$^{-1}$, depending on the window; halving the pulse gives the same increase
(Fig.~\ref{fig:pulse}d). This difference of growth rates measures the damping of the wave by the radiative response.

A radial displacement shifts the radiative-loss field across its mean gradient. In the linearized enthalpy equation,
the radiative response of the gas is $q_\xi\hat{\xi}+q_h\hat{h}$, where $q_\xi$ and $q_h$ are the derivatives of the
sink $q_\mathrm{rad}$ with respect to the mixture fraction and the enthalpy. For a radial displacement $\zeta$ of the
base fields, $\hat{\xi}=-\zeta\,\partial\xi/\partial r$ and $\hat{h}=-\zeta\,\partial h/\partial r$, and the response
becomes the steady sink displaced with the flame,
\begin{equation}
\hat{q}_\mathrm{rad}=q_\xi\,\hat{\xi}+q_h\,\hat{h}\approx-\zeta\,\frac{\partial q_\mathrm{rad}}{\partial r} .
\label{eq:qhat}
\end{equation}
We call the damping that this term exerts on the wave displaced-emission damping. On the outer flank of the flame, temperature and water vapor, and with them the emission, fall off outward. Gas
pushed outward therefore radiates more than the gas it replaces, and gas drawn inward radiates less, just as internal gravity waves in a stratified fluid lose their buoyancy anomaly by Newtonian cooling.

\begin{figure*}[!tbp]
\centering
\figfile{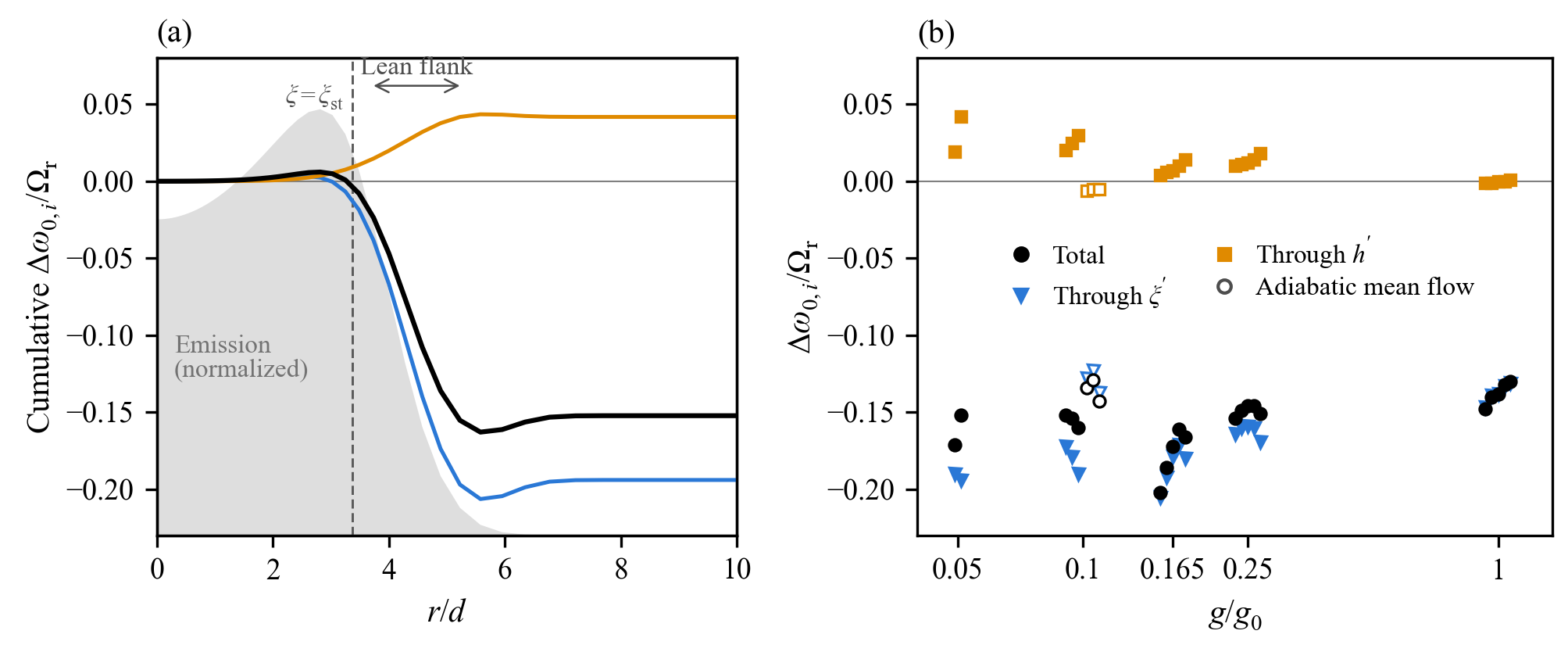}
\caption{Spatial origin and gravity dependence of displaced-emission damping in the complete linearization: (a) change
of the absolute growth rate, normalized by $\Omega_\mathrm{r}$, accumulated from the axis at $0.05\,g_0$ and $x/d=40$,
through $\xi'$ (blue), $h'$ (orange), and in total (black) (gray: emission; dashed: stoichiometric surface; arrow: lean flank); (b) the integrated contributions through $\xi'$ and $h'$, and their total at 23 stations of six flows (open: adiabatic mean
flow at $0.1\,g_0$).}
\label{fig:damploc}
\end{figure*}

The complete linearization (Section~\ref{sec:lsa}) quantifies this damping on the time-averaged flows. The radiative
response lowers the absolute growth rate by 0.13--0.19 times the radiative cooling rate of the products,
$\Omega_\mathrm{r}=\max\left(q_\mathrm{rad}/c_pT\right)$ (2.3--3.1\,s$^{-1}$ in these flames), without a trend with
gravity (Fig.~\ref{fig:damploc}b). The enthalpy component of the adjoint mode measures how sensitive the mode is to a local heat sink. Projected onto
this component, the displaced sink of Eq.~\eqref{eq:qhat} accounts for 95--99\% of the damping. The damping acts on the lean flank just outside the stoichiometric surface. At $0.05\,g_0$ and $x/d=40$, the cumulative
damping rises from 10 to 90\% of its total between $r/d=3.7$ and 5.2, while the stoichiometric surface lies at $r/d=3.4$
(Fig.~\ref{fig:damploc}a). There, the damping comes from the displaced mixture fraction; the part carried by the enthalpy is close to zero at
normal gravity and opposes the damping at reduced gravity, by up to a fifth of the mixture-fraction part (Fig.~\ref{fig:damploc}b). The
local contributions differ in phase and partly cancel, leaving a third to a half of their summed magnitude
(supplementary material): the damping depends on how the displaced sink and the adjoint enthalpy overlap in space and in phase, not on the size of
the emission alone. The same term lowers the growth rate at the
trailing-edge saddle of the impulse response by 0.41--0.51\,s$^{-1}$; these linear rates are lower than the rates measured with the pulses,
by up to a factor of two at $x/d=20$--40 (Fig.~\ref{fig:pulse}d).

\begin{figure*}[!tbp]
\centering
\figfile{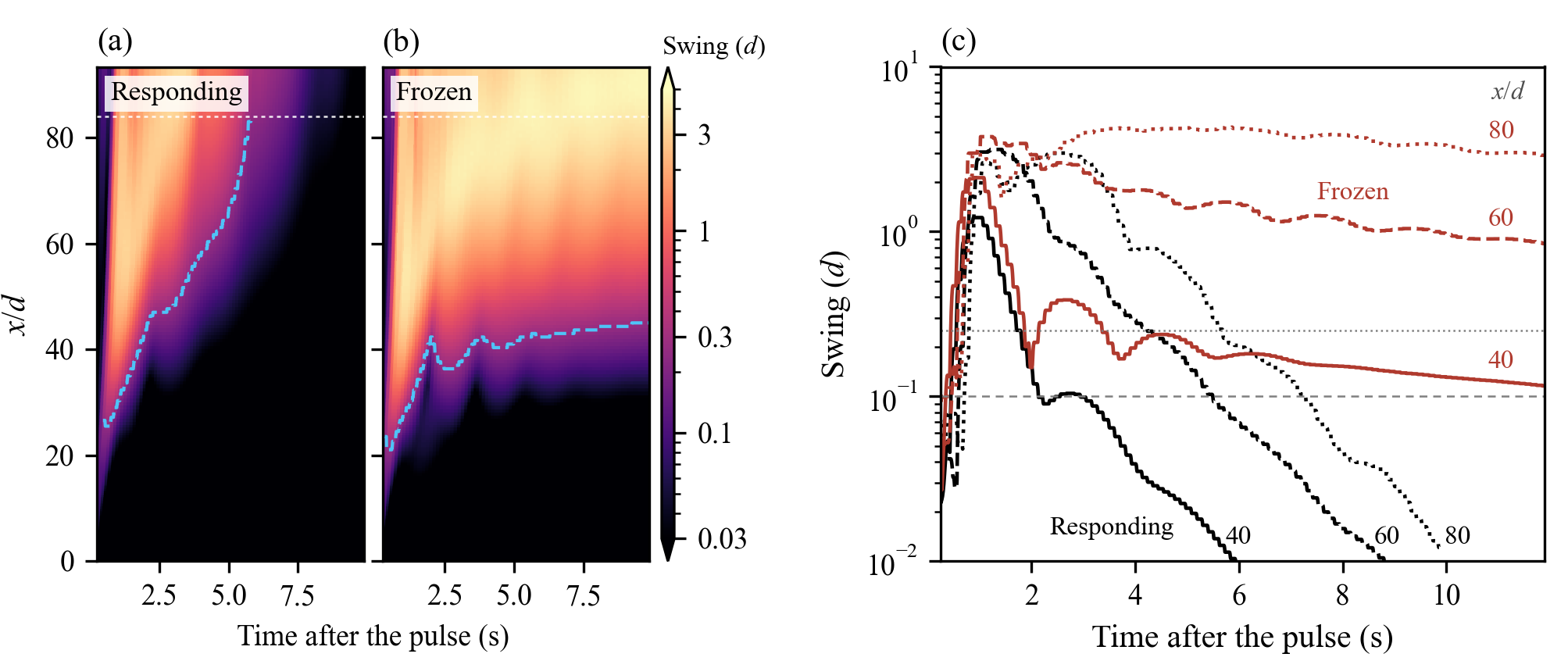}
\caption{Pulse of $0.025\,g_0$ on the steady radiating flame at $0.1\,g_0$: swing of the flame edge with the
radiative source (a) responding and (b) frozen (cyan dashed: onset; white dashed: $x/d=84$); (c) swing at $x/d=40$
(solid), 60 (dashed), and 80 (dotted), responding in black and frozen in red, with the thresholds $0.25d$ of the
onset and $0.1d$ of a lost oscillation (gray).}
\label{fig:frozen01}
\end{figure*}

Its consequence is clearest at $0.1\,g_0$, where the radiating flame has just lost its flicker
(Fig.~\ref{fig:frozen01}). There, a pulse of $0.025\,g_0$ excites a wave packet; comparing the frozen and responding flames while both are still
linear, as at $0.05\,g_0$, gives a damping rate of 0.6--0.9\,s$^{-1}$. The oscillation of the responding flame then recedes at about 9 diameters per second, as after the emission is
switched on in the adiabatic flame, and is lost from 7.9\,s on. With the source frozen, the
oscillation recedes as fast while the packet passes, but then slows down. From 3.6 to 5.6\,s, its onset recedes at 1.0 diameter per second, against 10.7 with the source responding. Over
the last 6\,s of the 12.1\,s record, it recedes at 0.5--0.8 diameters per second, ever more slowly. The record thus
shows a flame near its limit rather than a settled limit cycle (supplementary material). Without the pulse, the flame stays steady. The response of the emission to the disturbances thus sets how fast the flame loses its flicker; without it, the
flame stays near its limit.

Neither radiative rate falls with gravity, whereas the buoyant rates of the flicker do. The rate of the buoyancy loss,
$\gamma$, is the same at four gravity levels; the damping rate measured with the pulses lies in the same range at 0.05
and $0.1\,g_0$, and its linear counterpart shows no trend with gravity. With $\Omega_\mathrm{r}\approx2.7$\,s$^{-1}$,
both measured rates are a quarter to a third of $\Omega_\mathrm{r}$. Both radiative effects therefore become more important relative to the buoyant drive as gravity decreases. The
fractional loss of buoyancy, $\gamma t_\mathrm{res}$, rises from 0.04 at normal gravity to 0.4 at $0.05\,g_0$, and the
ratio of the radiative damping rate to the growth rate of the wave rises from 0.05 to 0.3 as that growth rate falls from about 20 to 3\,s$^{-1}$.

\subsection{Upstream spreading of disturbances against the co-flow}
\label{sec:coflow}\label{sec:interp}
Both radiative mechanisms lower the speed at which disturbances spread upstream through the hot layer relative
to the co-flow, $S=\Uco-V_-$ (Section~\ref{sec:lsa}). Where $S$ is smaller than the co-flow velocity, the co-flow sweeps disturbances downstream. The
oscillation therefore begins where $S$, rising along the layer, approaches $\Uco$, and near the limit, its upstream end
recedes at the speed $\Uco-S$ of the linear trailing edge. The limit itself follows the largest spreading speed,
$S_\mathrm{max}$, which scales with $(gd)^{1/2}$ without radiation; radiation lowers it by an amount that does not
scale with gravity.

The co-flow sets where the flicker begins and whether it persists, even without radiation. At normal gravity, a
stronger co-flow displaces the oscillation rather than suppressing it uniformly (Fig.~\ref{fig:onset}a; values in the
supplementary material): it raises the frequency, reduces the swing near the burner, and moves the oscillation
downstream, as observed in co-flowing methane flames~\cite{GohariDarabkhani2011,Fujisawa2016}. At 0.35\,m/s, the flame
edge near the burner is almost steady, while further downstream it swings more than at 0.15\,m/s, and at 0.7 and
0.8\,m/s ($\Frco=10.0$ and 13.0), the onset settles at $x/d=27.2$ and 35.7--37.7. A measurement confined to the near
field would therefore report a suppressed flicker, while the flame still oscillates further downstream, consistent with
the shift of the outer vortices beyond the visible flame observed at large co-flow
velocities~\cite{GohariDarabkhani2011}. On a shorter domain of $62d\times10d$, started from the flame at 0.5\,m/s, the flames at 0.6 and 0.7\,m/s lose their oscillation through the outflow boundary; these two outcomes are domain
effects and are not used to locate the limit. At reduced gravity, halving the co-flow can restore the flicker of the radiating flame. At $0.1\,g_0$, the halved
co-flow of 0.075\,m/s gives a sustained oscillation from $x/d=10.5$ once the slower co-flow has crossed the flame twice; at
$0.05\,g_0$, the oscillation that grows after the change recedes beyond $x/d=84$
(Fig.~\ref{fig:spacetime}c, f; Table~\ref{tab:lowg}).

If the velocity profile is fixed relative to the co-flow, changing the co-flow adds the same velocity to every ray
velocity of the local impulse response in the laboratory frame, so that $S=\Uco-V_-$ does not change: the spreading
speed is a property of the layer relative to the co-flow. In a simulation, changing the co-flow also changes the velocity excess
and the profile of the layer, which the analysis therefore takes from each time-averaged flow. The dotted lines in Fig.~\ref{fig:onset}d mark $S=\Uco$,
above which the layer is absolutely unstable. In the flame at normal gravity and 0.7\,m/s, $S$ rises along the layer
to 0.84 times the co-flow velocity at the onset; at $0.25\,g_0$ and 0.35\,m/s, it stays 13\% below the co-flow
velocity, and the oscillation of this flame recedes slowly. In the steady flames at $0.05\,g_0$, $S$ reaches 0.107\,m/s at $x/d=60$ without radiation and is still rising there;
with radiation, it peaks at only 0.066\,m/s near $x/d=30$ and then falls as the radiating layer cools. Beyond
$x/d=60$, the co-flow outside these flames is no longer uniform.

The local analysis places the onset where $S$ overtakes $\Uco$, as in methane flames at normal
gravity~\cite{Lingens1996,LingensReeker1996} (Fig.~\ref{fig:onset}c). Of the thirteen oscillating flames analyzed, ten
are absolutely unstable in the axisymmetric mode of the hot layer. In all ten, the upstream boundary of the absolutely
unstable region lies within $3.7d$ of the onset defined with $0.1d$, although the boundary itself ranges from $x/d=0.7$ to 32 with co-flow, gravity, and radiation. At this boundary, the absolute frequency, which selects the
frequency of nonlinear global modes~\cite{Pier2002,Lesshafft2006}, lies 10--20\% below the simulated frequency in six of them (at normal gravity with 0.15, 0.35, and 0.5\,m/s; the
radiating flames at 0.165 and $0.25\,g_0$; the adiabatic flame at $0.165\,g_0$; supplementary material). Two adiabatic
flames are marginal, and in the flame at normal gravity and 0.7\,m/s, the closest to its limit, the layer is
convectively unstable at every station upstream of the onset.

In the simulations, the onset follows the velocity of the hot layer relative to the co-flow. In the sustained flames at
normal gravity with co-flows of 0.15--0.5\,m/s and at reduced gravity with 0.15\,m/s, the oscillation begins where the
velocity excess of this layer over the co-flow, $\Delta U$, is 4.2--6.3 times the
co-flow velocity (swing $0.25d$; 3.6--4.9 for $0.1d$), while the onset position ranges from $x/d=3$ to 22. Unlike in light
jets~\cite{JendoubiStrykowski1994,Hallberg2007}, the ratio of the co-flow to the centerline jet velocity, 2--9\%, does not distinguish the flames that keep their
flicker from those that lose it. Without radiation and with unity Lewis numbers, the onsets obtained by reducing gravity at
fixed co-flow and by increasing the co-flow at normal gravity therefore fall on one line,
$x_\mathrm{on}/d\approx2+3.2\,\Frco$, to within about 20\% up to $\Frco=13$ (Fig.~\ref{fig:onset}b), whereas their
frequencies are not ordered by $\Frco$ (supplementary material). The line follows from the buoyant acceleration of the layer. With $U_\mathrm{hot}^2-\Uco^2\approx2g\beta_\mathrm{eff}x$,
the velocity excess reaches $\Lambda\Uco$ at $x/d=\Lambda(\Lambda+2)\Frco/(2\beta_\mathrm{eff})$. At the onsets,
$\Lambda=4.2$--6.3, so that
the slope 3.2 implies an effective buoyancy factor $\beta_\mathrm{eff}=4$--8, against
$\beta=\rho_\mathrm{air}/\rho_\mathrm{hot}-1\approx8$ at the temperature peak.

\begin{figure*}[!tbp]
\centering
\figfile{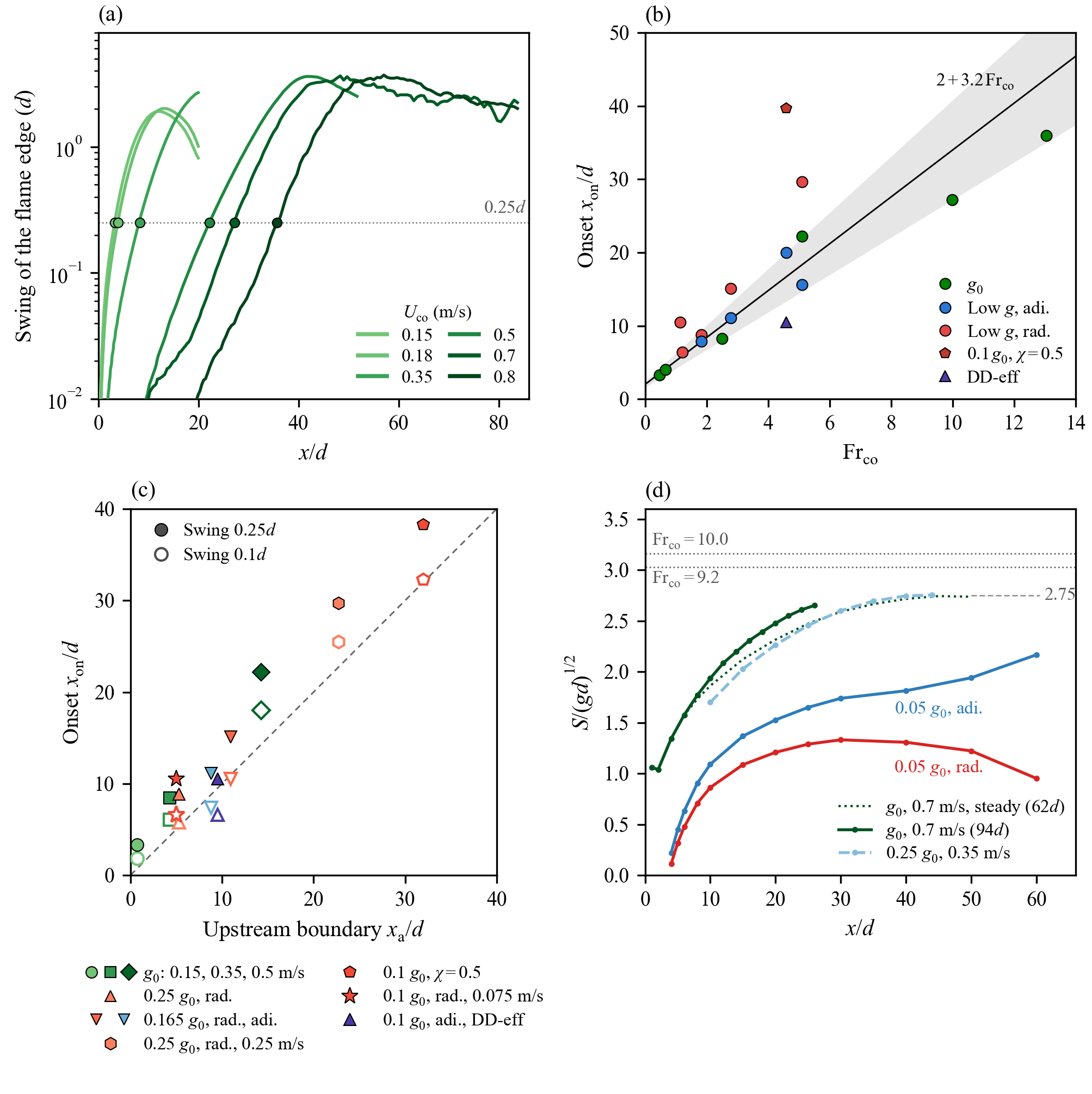}
\caption{Onset of the flicker: (a) swing of the flame edge at normal gravity, each co-flow on its longest domain
(dots: onset); (b) onset against $\Frco$ with the line $x_\mathrm{on}/d=2+3.2\,\Frco$ (band: $\pm20\%$, for comparison; green:
normal gravity; blue and red: adiabatic and radiating flames at reduced gravity; pentagon: $\chi=0.5$; triangle:
DD-eff); (c) onset against the upstream boundary $x_\mathrm{a}$ of absolute instability (open: swing $0.1d$);
(d) upstream spreading speed $S=\Uco-V_-$ scaled by $(gd)^{1/2}$ (dotted: $\Frco^{1/2}$).}
\label{fig:onset}
\end{figure*}

Where the layer is convectively unstable, $\Uco-S$ is the velocity at which the trailing edge of a disturbance is
swept downstream, and the upstream end of a saturated oscillation, its trailing front, recedes like a pulled front, at the speed of the linear trailing
edge~\cite{vanSaarloos2003}. The emission series at $0.1\,g_0$ tests this. With 0.70, 0.75, and 0.80 of the emission, the swing stays at 1.2--4$d$, and $\Uco-S$ at the onset is 1.7, 2.6, and
3.7 diameters per second. Defining the onset with a swing of 0.1, 0.25, or $0.5d$ gives a range of recession speeds for each
flame: 1.7--2.4 diameters per second for $\chi=0.70$, 2.0--3.4 for $\chi=0.75$, and 3.0--5.0 for $\chi=0.80$. Each range
contains the corresponding $\Uco-S$ (Fig.~\ref{fig:dose}d). Near the limit,
where the swing stays nearly constant, the linear trailing edge thus predicts how fast the oscillation recedes. With the full emission, the
oscillation decays as it recedes, its largest swing falling from 3.2 to $0.9d$; contours of fixed swing then move at $-\partial_tA/\partial_xA$, faster than the linear trailing edge: 7--10 against 3.3--6.7 diameters per second. Whether the
oscillation decays depends on displaced-emission damping, which the simplified hydrodynamic model leaves out.

Without radiation, the largest spreading speed scales with $(gd)^{1/2}$: $S_\mathrm{max}/(gd)^{1/2}=2.74$ at $g_0$ and
2.75 at $0.25\,g_0$, a factor of four apart in gravity (Fig.~\ref{fig:onset}d). This scaling is why the onsets are
ordered by the co-flow Froude number, and it sets the buoyant part of the limit of the flicker. The local criterion
$S_\mathrm{max}=\Uco$ places this limit at $\Frco\approx7.5$, below the limits reached in the simulations.

\subsection{A radiation-modified criterion for sustained flicker}
\label{sec:limit}
\begin{figure*}[!tbp]
\centering
\figfile{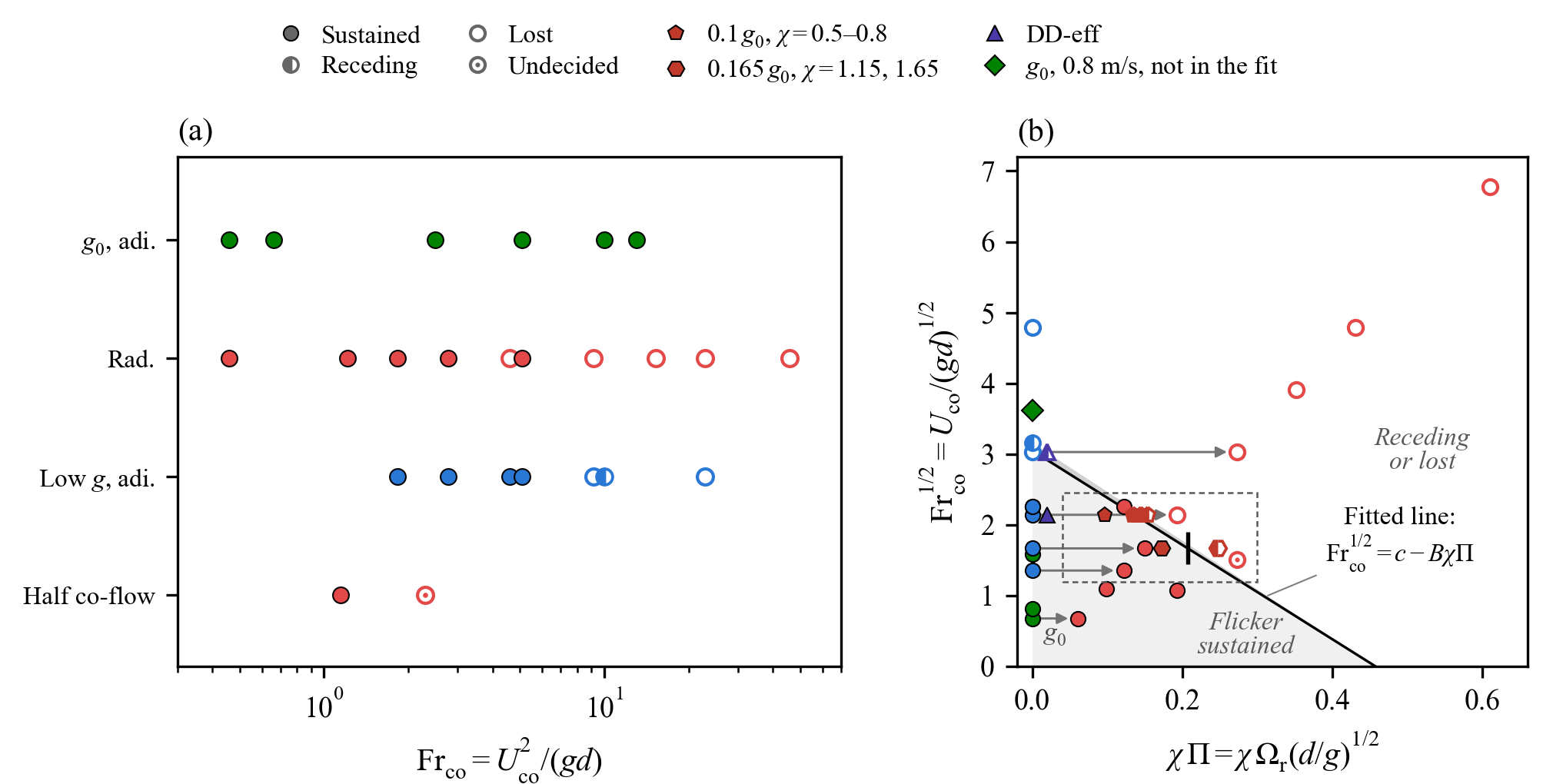}
\caption{Limit of the flicker: (a) outcome against $\Frco$; (b) all cases in the plane of $\chi\Pi$ and
$\Frco^{1/2}$ with the line of Eq.~\eqref{eq:pi} ($c=3.05$, $b=0.09$\,m/s; band: $U_\mathrm{c}/\Uco=0.965$--1.007, where
sustained and other cases overlap); arrows: optically thin emission at $\Uco=0.15$\,m/s from $g_0$ (bottom) to $0.05\,g_0$ (top); bar:
limiting emission at $0.165\,g_0$ predicted a priori; dashed box: region of Fig.~\ref{fig:test}a.}
\label{fig:state}
\end{figure*}

Without radiation, the limit of the flicker follows the co-flow Froude number. Every adiabatic flame with $\Frco\le5.1$ sustains its oscillation, and at normal gravity the
oscillation persists up to at least $\Frco=13.0$; it recedes slowly at $0.25\,g_0$ with $\Frco=10.0$ and is lost at $0.05\,g_0$ with $\Frco=9.2$
(Fig.~\ref{fig:state}a). The adiabatic limit thus lies near
$\Frco\approx9$--10 at reduced gravity and above 13 at normal gravity, so that the critical co-flow, the largest at which the
flicker persists, falls at least as fast as $g^{1/2}$ between normal gravity and $0.25\,g_0$. Extrapolated to this jet, the correlation of Fujisawa et
al.~\cite{Fujisawa2016} for methane flames gives $\Frco=9.6$ at normal gravity, where the present flame keeps its
flicker at 13.0, and a critical co-flow that falls only as $g^{0.37}$. The methane cup-burner flame of
Ref.~\cite{KattaTakahashiLinteris2004} has its cut-off two orders of magnitude lower, at $\Frco=0.04$--0.08.

Differential diffusion moves the onset upstream but leaves this limit within the same gravity interval. With DD-eff, the adiabatic flame at $0.1\,g_0$ keeps its oscillation, from $x/d=10.5$ instead of 20.0 and at a
frequency 19\% lower. At $0.05\,g_0$, where the flame with unity Lewis numbers loses its flicker, the oscillation of
the DD-eff flame recedes: its onset moves downstream from $x/d=33.7$ to 37.0 over the last 2\,s of an 8.4\,s record
(Table~\ref{tab:lowg}). With both closures, the adiabatic flame thus loses its sustained flicker between
0.05 and $0.1\,g_0$.

Radiating flames are not ordered by $\Frco$ alone (Fig.~\ref{fig:state}a). At the co-flow of the experiment, they keep
the flicker up to $\Frco=2.8$ and lose it at 4.6, whereas at $0.25\,g_0$, a radiating flame keeps it at 5.1; with the
co-flow halved, the flicker returns at $\Frco=1.15$ but recedes at 2.29. A second parameter is needed: the decrement
of the spreading speed caused by radiation, to which both radiative mechanisms contribute.

The loss of buoyancy reduces the hot-layer velocity. In a layer accelerated by buoyancy from the co-flow
velocity, $U_\mathrm{hot}^2-\Uco^2\approx2g\int\beta\,\mathrm{d}x$; for uniform $\beta$ and a weak loss of buoyancy $\gamma\chi t_\mathrm{res}$ following Eq.~\eqref{eq:reslaw}, with
$t_\mathrm{res}=\int\mathrm{d}x/U_\mathrm{hot}$, the layer velocity falls by
\begin{equation}
\delta U_\mathrm{hot}\approx\gamma\chi x\,\frac{(U_\mathrm{hot}-\Uco)(2U_\mathrm{hot}+\Uco)}{3U_\mathrm{hot}(U_\mathrm{hot}+\Uco)} .
\label{eq:du}
\end{equation}
At fixed $x$ and velocity ratio $U_\mathrm{hot}/\Uco$, Eq.~\eqref{eq:du} does not depend on gravity explicitly: in the
buoyancy-dominated limit, the residence time at a given station falls as $g^{-1/2}$, while the velocity rises as
$g^{1/2}$, and the two dependences cancel. The velocity-ratio factor in Eq.~\eqref{eq:du} is 0.49--0.54 at the onsets and tends to 2/3 for a negligible co-flow. With $S=s\,\Delta U$, the spreading speed then falls by $b\chi$, where $b\approx0.5\,s\,\gamma\,x_*$; here
$x_*\approx0.2$--0.3\,m is the position of $S_\mathrm{max}$ and $s=S/\Delta U=0.14$--0.32. To first order in a weak,
uniform loss, $b\approx0.01$--0.04\,m/s, of the order of the
decrement that the local analysis gives at $0.05\,g_0$ (0.02--0.06\,m/s between $x/d=30$ and 60).

Displaced-emission damping directly lowers the spreading speed. On a fixed mean flow, a damping that lowers the growth
rate at the saddle point $k_-$ of the trailing edge by $\delta\sigma>0$ moves the trailing edge downstream. To first order~\cite{vanSaarloos2003}, the corresponding change of the spreading speed is
\begin{equation}
\delta S\approx-\frac{\delta\sigma}{|\mathrm{Im}\,k_-|} .
\label{eq:ds}
\end{equation}
In the complete linearization on fixed mean flows, the radiative response lowers $S$ in proportion to its strength, by
0.009\,m/s at $0.25\,g_0$ and 0.012\,m/s at $0.05\,g_0$ at full strength (five stations). With the damping at the
trailing-edge saddle, $\delta\sigma=0.41$--0.51\,s$^{-1}$, Eq.~\eqref{eq:ds} reproduces these values to within 5\%
(supplementary material). This decrement rises by a third, while $(gd)^{1/2}$ changes by a factor of 2.2.

Neither decrement scales with gravity, whereas the buoyant spreading speed scales with $(gd)^{1/2}$. The critical co-flow is therefore
\begin{equation}
U_\mathrm{c}=c\,(gd)^{1/2}-b\,\chi ,
\label{eq:uc}
\end{equation}
with a buoyant constant $c$, about 2.75 in the local analysis, and a radiative constant $b$. Divided by $(gd)^{1/2}$,
Eq.~\eqref{eq:uc} becomes a relation between two dimensionless numbers,
\begin{equation}
\mathrm{Fr}_\mathrm{co,c}^{1/2}=c-B\,\chi\,\Pi,\qquad \Pi=\Omega_\mathrm{r}\left(\frac{d}{g}\right)^{1/2},\qquad
B=\frac{b}{\Omega_\mathrm{r}d}\approx6.7 ,
\label{eq:pi}
\end{equation}
where the radiation number $\Pi$ compares the buoyant time $(d/g)^{1/2}$ with the radiative time scale $1/\Omega_\mathrm{r}$ of the products ($\Omega_\mathrm{r}\approx2.7$\,s$^{-1}$). It is a
radiation Damk\"ohler number built on the buoyant time, of the kind that Qin et al.~\cite{Qin2004} related to the
Froude number in partially premixed flames; here it enters the limit of the flicker. In Eq.~\eqref{eq:pi}, $c-\Frco^{1/2}$ is the buoyant margin of the hot layer against the co-flow, and $B\chi\Pi$ is the
part of this margin that radiation removes; the flicker persists while $B\chi\Pi<c-\Frco^{1/2}$. If $b$ grows with the length of the hot layer, as the first-order estimate $b\approx0.5\,s\,\gamma\,x_*$
suggests, $B$ is the same for geometrically similar flames burning the same mixture, whereas at fixed gravity, $\Pi$
grows as $d^{1/2}$: radiation then weighs more on larger flames.

The derivation fixes the form of the relation, not the size of its constants. The constants are chosen so that $U_\mathrm{c}/\Uco\ge1$ for the flames with a sustained
oscillation and $U_\mathrm{c}/\Uco\le1$ for the lost and receding ones. The calibration uses the 27 cases with unity Lewis numbers computed before the test below, at reduced gravity and at
normal gravity up to 0.7\,m/s. With $c=3.05$ and $b=0.09$\,m/s, the sustained cases lie at $U_\mathrm{c}/\Uco\ge0.965$
and the lost and receding ones at $U_\mathrm{c}/\Uco\le1.007$, a tolerance of 3.5\%. The admissible $(c,b)$ pairs,
which meet the same tolerance, lie within $c=3.05$--3.13 and $b=0.084$--0.106\,m/s (supplementary material). In the plane of $\chi\Pi$ and $\Frco^{1/2}$, the outcomes at reduced
gravity thus fall on either side of a straight line, which divides it into a region of sustained flicker and one where
the flicker recedes or is lost (Fig.~\ref{fig:state}b).

\begin{figure*}[!tbp]
\centering
\figfile{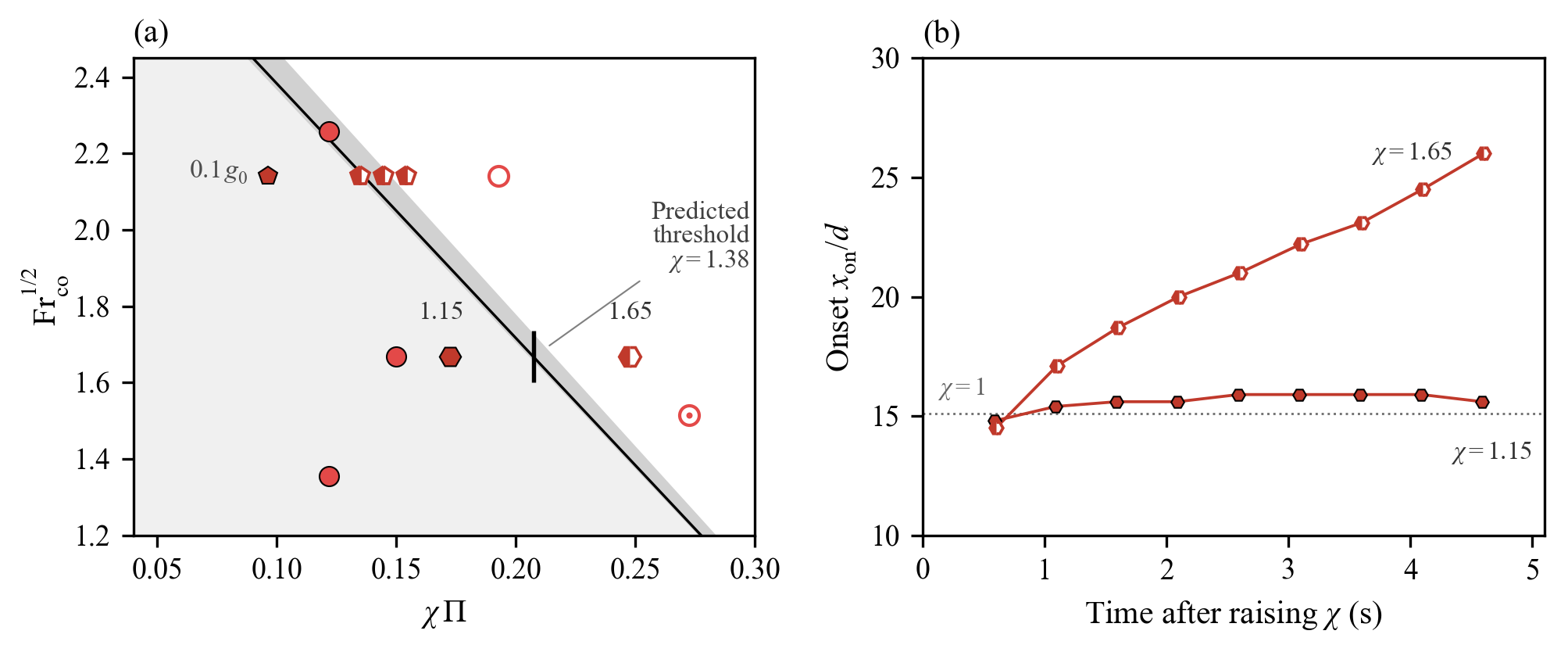}
\caption{A-priori test of the limit at $0.165\,g_0$: (a) region of the dashed box of Fig.~\ref{fig:state}b, with the
limiting emission predicted before the runs (bar, $\chi=1.38$) and the two emission factors computed afterwards
(hexagons); (b) onset after the emission is raised from $\chi=1$ (dotted) to 1.15 and 1.65.}
\label{fig:test}
\end{figure*}

Before the two test flames were computed, the calibrated relation placed the limiting emission at $0.165\,g_0$ at
$\chi=1.38$; the flame with $\chi=1$ at this gravity belongs to the calibration. Two new emission factors, 1.15 and
1.65, were chosen so that the flame at $0.165\,g_0$ has the same $U_\mathrm{c}/\Uco$ as the flames with $\chi=0.5$
and 1 at $0.1\,g_0$. Both fall on the predicted sides (Fig.~\ref{fig:test}): with 1.15, the flame keeps its flicker, its onset fixed at
$x/d=15.6$--15.9, and with 1.65, its oscillation recedes at 2.5 diameters per second. Both outcomes agree with the prediction for every admissible pair; across these pairs, the predicted limiting emission
at this gravity ranges from $\chi=1.25$ to 1.49. With the fitted $c$, the two flames bound $b$ at this gravity to 0.075--0.108\,m/s, around the fitted 0.09\,m/s.

The calibrated $b$ exceeds the linear estimates of the two channels, even taken together (Table~\ref{tab:budget}).
The frozen flame at $0.1\,g_0$ indicates how $b$ divides between them. Its hot layer is that of the radiating flame to
within 0.1\% upstream of the onset (supplementary material), and it recedes at a speed between those of the flames
with $\chi=0.5$ and 0.7. Assume that it shares the critical co-flow of a responding flame with $0.5<\chi<0.7$ and that the two channels
scale with the emission and add. The loss of buoyancy then supplies about two thirds of $b$, and displaced-emission
damping the remaining third.

\begin{table*}[!tbp]
\centering\footnotesize
\caption{Estimates of the radiative decrement $b$ of the critical co-flow, Eq.~\eqref{eq:uc}, and of its two channels,
from different models, in m/s; the frozen-source shares rest on the assumptions stated in the text.}
\label{tab:budget}
\begin{tabular}{lccc}
\toprule
Estimate & \shortstack{Loss of\\buoyancy} & \shortstack{Displaced-emission\\damping} & $b$ \\
\midrule
First order, Eq.~\eqref{eq:du} & 0.01--0.04 & -- & -- \\
Local analysis, $0.05\,g_0$ & 0.02--0.06 & -- & -- \\
Complete linearization, Eq.~\eqref{eq:ds} & -- & 0.009--0.012 & -- \\
Frozen source, shares of $b$ & 0.045--0.063 & 0.027--0.045 & -- \\
Calibrated, $b=0.09$ & -- & -- & 0.084--0.106 \\
Two flames at $0.165\,g_0$ & -- & -- & 0.075--0.108 \\
\bottomrule
\end{tabular}
\end{table*}

The local analysis thus establishes the two channels and the scaling of the limit with gravity, while the constants
of the limit are calibrated against the simulations; for the buoyant constant, too, the local criterion gives
$c\approx2.75$ against the calibrated 3.05. At normal gravity, the flame at 0.8\,m/s keeps its flicker at
$\Frco=13.0$, which requires $c\ge3.6$ there. The relation is therefore proposed for reduced gravity, where the
radiative term is comparable with the buoyant margin $c-\Frco^{1/2}$, whereas at normal gravity, radiation takes only a
sixth of that margin. Since both radiative mechanisms act through the
emission and the residence time of the products, differential diffusion is expected to shift the constants of the
radiating limit rather than its form; a radiating flame with differential diffusion, which requires an enthalpy transport that retains the differential-diffusion terms, would test this.

The relation shows how much of the buoyant margin radiation consumes. At the co-flow of the experiment, the adiabatic
flame has a margin $c-\Frco^{1/2}$ of 0.91 at $0.1\,g_0$ and 0.02 at
$0.05\,g_0$, while the radiative term $B\chi\Pi$ of the optically thin emission is 1.29 and 1.82. Radiation thus takes 1.4
times the margin at $0.1\,g_0$ and overtakes it below $0.126\,g_0$, where $\Pi\approx0.17$ (0.12--0.14\,$g_0$ for every admissible pair). The flicker ends below the gravity $g_\mathrm{c}=(\Uco+b\chi)^2/(c^2d)$. Independently of $c$, radiation with the fitted $b$
raises it by the factor $[(\Uco+b\chi)/\Uco]^2\approx2.6$, from 0.05 to 0.13 times normal gravity at the co-flow of the
experiment; with the net radiative loss estimated with reabsorption, $\chi\approx0.72$--0.80, the factor is still 2.0--2.2,
and the critical gravity 0.10--0.11 times normal gravity.

\subsection{Transient flicker and finite observation times}
\label{sec:transient}
Beyond its limit, a flame does not lose its oscillation at once. The oscillation that follows a reduction of gravity
is a packet of the kind that the gravity pulses excite, and the flame loses it first near the burner and then
progressively farther downstream. Near the burner ($x/d\le30$, about the length of the flame at normal gravity), the flames are steady within
1.0--1.8\,s of the step, whereas further downstream the packet keeps the flame oscillating for seconds
(Fig.~\ref{fig:spacetime}a, d). At $0.05\,g_0$, the packet needs about 2.5\,s to leave a flame about 0.45\,m long. While it
passes, the packet oscillates at a frequency close to that of a sustained flicker at the same gravity, 4.53\,Hz at
$0.1\,g_0$ with radiation (Table~\ref{tab:lowg}): a familiar flicker frequency does not show that the flame sustains
the flicker.

Three questions must therefore be kept apart: whether the reduced-gravity flame sustains the flicker, how long the
packet needs to leave the flame, and how long the observation lasts. Drop towers provide 2.2--9.3\,s of
microgravity~\cite{Bahadori1990,vonKampen2006,Konemann2015}. A short drop-tower record of the whole flame can therefore capture a transient wave packet even when the flame does
not sustain flicker, whereas an observation confined to the near field would show a steady flame. Reports of
reduced-gravity flicker from simulations are likewise difficult to interpret without the length of the
record~\cite{RoquemoreKatta2000}. In a 2.2\,s drop tower, the surface of a hydrogen jet flame became steady, while the
surrounding flow was still evolving at the end of the drop~\cite{AlAmmar2000}. The transient times computed here belong
to this flame and to a step reduction of gravity, but the distinction between convected wave packets and sustained
flicker applies to any short reduced-gravity record.

%=====================================================================================================================
\section{Conclusions}
\label{sec:conclusions}
Gas radiation raises the gravity below which a co-flowing hydrogen jet diffusion flame can no longer sustain
flicker. In the simulations, the radiating flame loses its flicker between 0.1 and $0.165\,g_0$, the adiabatic flame only between 0.05 and $0.1\,g_0$. At the co-flow of the experiment, the calibrated criterion
places the critical gravity at 0.13 instead of 0.05 times normal gravity, a 2.6-fold increase, and the increase is at
least twofold with the net radiative loss estimated with reabsorption. At one tenth of
normal gravity, the emission alone decides whether the flame flickers, and this net loss still removes the flicker. A loss that
barely changes the flicker at normal gravity thus controls it at reduced gravity.

Radiation acts through two mechanisms whose rates do not fall with gravity. It removes buoyancy from the hot outer
layer in proportion to the residence time of the gas, and it damps the disturbances themselves on the lean flank of
the flame, where the emission moves with the gas that the wave displaces. This displaced-emission damping is the gas-radiation
counterpart of the Newtonian cooling of internal gravity waves. Freezing the radiative source separates the two.
Retaining the mean radiative cooling while suppressing the response of the source reduces the speed at which the
oscillation recedes at one tenth of normal gravity to less than a tenth of that with the responding source, and the flame stays near the limit of its
flicker. Cooling of the
mean flow brings the flame close to its limit, and the damping of the disturbances carries it beyond.

Both mechanisms lower one quantity, the speed at which disturbances spread upstream through the hot layer against the
co-flow. The onset lies where this spreading balances the advection by the co-flow, and near the
limit, the oscillation recedes like a pulled front, at the speed of the linear trailing edge. Without radiation, the
spreading speed scales with $(gd)^{1/2}$, which orders the onsets and the adiabatic limits by the co-flow Froude number;
radiation lowers it by an amount that does not scale with gravity. At reduced gravity, the limit of the flicker is
therefore the straight line $\mathrm{Fr}_\mathrm{co,c}^{1/2}=c-B\chi\Pi$, in which the radiation number $\Pi$ compares
the buoyant time with the radiative time scale of the products. The two scalings fix its form, and the simulations fix its constants. Once calibrated, the line predicted on which
side of the limit two new flames lie. The gravity at
which a co-flowing jet flame stops flickering is thus set by the co-flow and the radiative loss rather than by the flame alone, and for a high-velocity jet from a small nozzle, it need not lie near zero gravity. If
the radiative constant grows with the length of the hot layer, radiation also weighs more on larger flames.

Beyond its limit, a flame still amplifies disturbances into wave packets that take seconds to leave it and oscillate
at the familiar flicker frequency, so that an oscillation seen during the few seconds of a drop-tower test need not
be flicker that the reduced-gravity flame sustains. These conclusions rest on one burner and jet velocity, unity Lewis
numbers, and an optically thin emission at frozen composition; with differential diffusion, the adiabatic limit stays
within the same gravity interval. Other nozzles and jet velocities, radiating flames with differential diffusion, and
reduced-gravity experiments with a controlled co-flow would test the constants of the relation. Where low gravity leaves the buoyant spreading of disturbances only slightly faster than the co-flow, gas radiation
decides whether the flame flickers.

\section*{Supplementary Material}
The supplementary material, appended after the references, contains the two further thermochemical closures, checks of the radiation model, the
numerical method and tests, details of the local stability analysis, the measured temperature histories, the wave
structure at normal gravity, the emission-factor series, the gravity pulses and the radiative damping, the co-flow
series, the calibration of the limit relation, and a list of symbols.

\begin{acknowledgments}
This research received no external funding. During the preparation of this work, the authors used Claude Code with
Claude Opus~5.5 (Anthropic) to assist with the \LaTeX{} editing and language polishing of the manuscript, to set up the cluster environment, and
to monitor the simulations. The authors reviewed and edited the output, and they take full responsibility for the content of this
publication.
\end{acknowledgments}

\section*{Author Declarations}
\subsection*{Conflict of Interest}
The authors have no conflicts to disclose.

\subsection*{Author Contributions}
\textbf{Tianfang Xie}: Conceptualization (lead); Data curation (lead); Formal analysis (lead); Investigation (lead);
Methodology (lead); Software (lead); Validation (lead); Visualization (lead); Writing -- original draft (lead);
Writing -- review \& editing (lead). \textbf{Meng Wang}: Data curation (supporting); Validation (supporting);
Visualization (supporting); Writing -- original draft (supporting); Writing -- review \& editing (supporting).

\section*{Data Availability}
The experimental data used for comparison are published in Ref.~\onlinecite{Dennis2016}. The simulation data that
support the findings of this study are available from the corresponding author upon reasonable request. The solver
used in this study is not publicly available.

\bibliography{refs}

\clearpage
\FloatBarrier
\setcounter{section}{0}\setcounter{figure}{0}\setcounter{table}{0}\setcounter{equation}{0}
\renewcommand{\thesection}{S\arabic{section}}\renewcommand{\theHsection}{SI.\arabic{section}}
\renewcommand{\thefigure}{S\arabic{figure}}\renewcommand{\theHfigure}{SI.\arabic{figure}}
\renewcommand{\thetable}{S\arabic{table}}\renewcommand{\theHtable}{SI.\arabic{table}}
\begin{center}
{\large\bfseries Supplementary material}\\[0.5ex]
Equations, figures and tables of the article are cited without the prefix S.
\end{center}

%=====================================================================================================================
\section{Thermochemical closures}
\label{SI-sec:closures}
\begin{figure}[!htbp]
\centering
\includegraphics[width=3.37in]{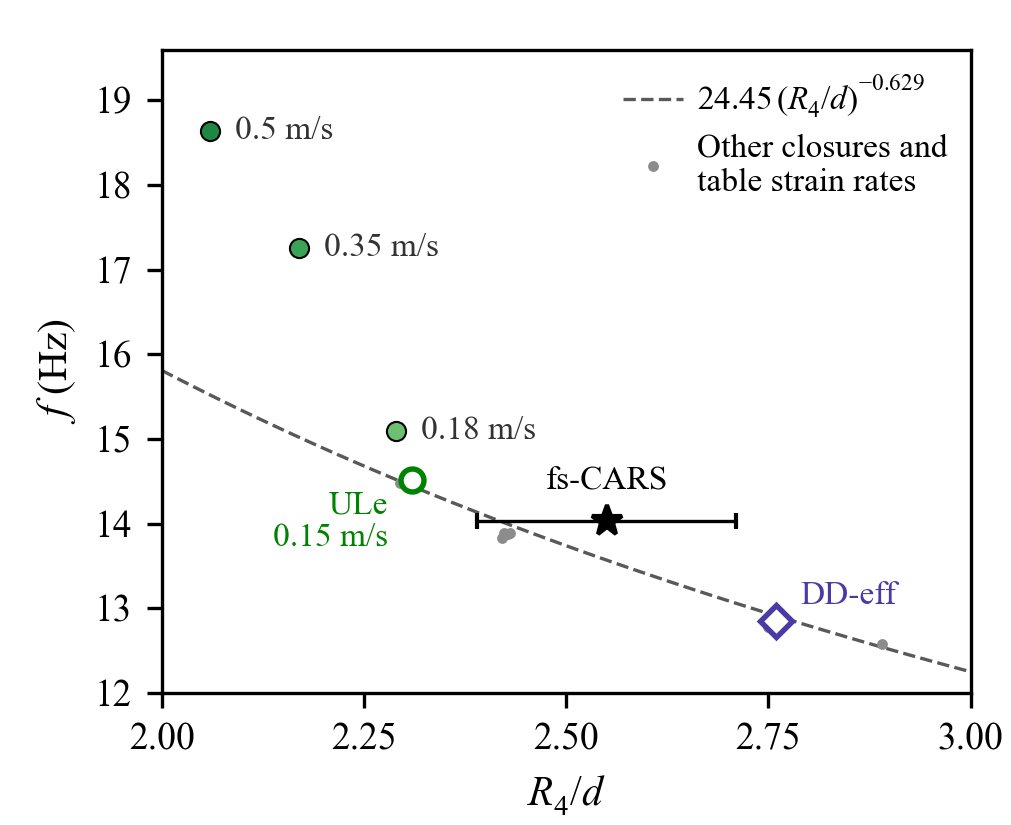}
\caption{Flicker frequency against the outer 1000\,K radius of the mean flame at $x/d=4$ at normal gravity: ULe and
DD-eff, the other closures and strain rates (gray), the ULe co-flow series (labels: $\Uco$), and the measurement;
dashed: fit to the four closures at 0.15\,m/s.}
\label{SI-fig:fR}
\alttext{Scatter plot of the flicker frequency against the outer 1000\,K flame radius at $x/d=4$ at normal gravity for four thermochemical closures and strain rates, the co-flow series, and the measurement; the frequency decreases with the radius, and the co-flow series lies above the closure trend.}
\end{figure}

Besides ULe and DD-eff (Section~II~B of the article), two closures differ in how the table coordinate is transported.
DD-$\alpha$ transports Bilger's mixture fraction $\xi_\mathrm{B}$ of the differential-diffusion table with the thermal
diffusivity $\alpha$ instead of its effective diffusivity at stoichiometry, $1.77\alpha$; DD-Z$\alpha$ tabulates the
differential-diffusion flamelets against a passive scalar that diffuses with $\alpha$. No closure matches all measured
quantities (Table~\ref{SI-tab:closures}). Across the closures and the strain rates of the table ($a=10$--100\,s$^{-1}$),
the frequency follows the flame radius, the closure changing it by up to 1.9\,Hz and the strain rate by less than
0.1\,Hz; the co-flow series lies above this trend (Fig.~\ref{SI-fig:fR}).

\begin{table}[!htbp]
\caption{The four closures against fs-CARS at normal gravity ($\Uco=0.15$\,m/s, $30d\times10d$ domain). Edge
swing: coherent (phase-averaged) peak-to-peak value of $R$; measured: inferred with the moving-edge model of
Section~\ref{SI-sec:cars}. Measured radius at $x/d=4$: mean of the two sides $\pm$ half their difference. Coherent p-p:
peak-to-peak temperature relative to the measurement (medians over the measurement points).}
\label{SI-tab:closures}
\begin{tabular}{lccccc}
\toprule
 & ULe & DD-$\alpha$ & DD-eff & DD-Z$\alpha$ & fs-CARS \\
\midrule
1000\,K radius at $x/d=1$ ($d$) & 1.66 & 1.74 & 1.96 & 2.06 & -- \\
\hspace{1em}at $x/d=4$ & 2.31 & 2.43 & 2.76 & 2.89 & $2.55\pm0.16$ \\
\hspace{1em}at $x/d=8$ & 2.86 & 3.02 & 3.52 & 3.74 & 2.65 \\
Peak mean $T$, $x/d\le5$ (K) & 2156--2186 & 2389--2426 & 2410--2430 & 2405--2429 & 2142--2343 \\
Frequency (Hz) & 14.51 & 13.89 & 12.85 & 12.58 & 14.03 \\
Edge swing at $x/d=4$ ($d$) & 0.35 & 0.38 & 0.48 & 0.51 & 0.19--0.21 \\
\hspace{1em}at $x/d=8$ & 1.25 & 1.29 & 1.57 & 1.66 & 0.51--0.55 \\
\hspace{1em}at $x/d=16$ & 1.48 & 1.61 & 2.26 & 2.47 & 1.24--1.37 \\
Coherent p-p / measured & 1.02 & 1.27 & 1.69 & 1.49 & 1 \\
\bottomrule
\end{tabular}
\alttext{Table comparing four thermochemical closures with the fs-CARS measurements at normal gravity: flame radius at three heights, peak mean temperature, flicker frequency, edge swing at three heights, and coherent temperature amplitude relative to the measurement.}
\end{table}

\FloatBarrier
%=====================================================================================================================
\section{Radiation model}
\label{SI-sec:radiation}
Compared with radiative counterflow flames with enthalpy losses of 0.8--1.0\,MJ/kg, the frozen-composition table
underestimates the temperature by 19--45\,K. In the steady radiating flame at $0.05\,g_0$, the radiated power and the
outflow of the defect balance to within 1.4\%. The reabsorption estimate of Section~II~C of the article uses 42 gray
gases. In the steady radiating flame at $0.1\,g_0$, it finds 16\% of the optically thin emission of the gas above
1000\,K reabsorbed and a Planck-mean coefficient about 10\% below that of the TNF model, so that the net loss of this
gas is 0.755 times the sink of the simulations.

\begin{figure}[!htbp]
\centering
\includegraphics[width=\textwidth]{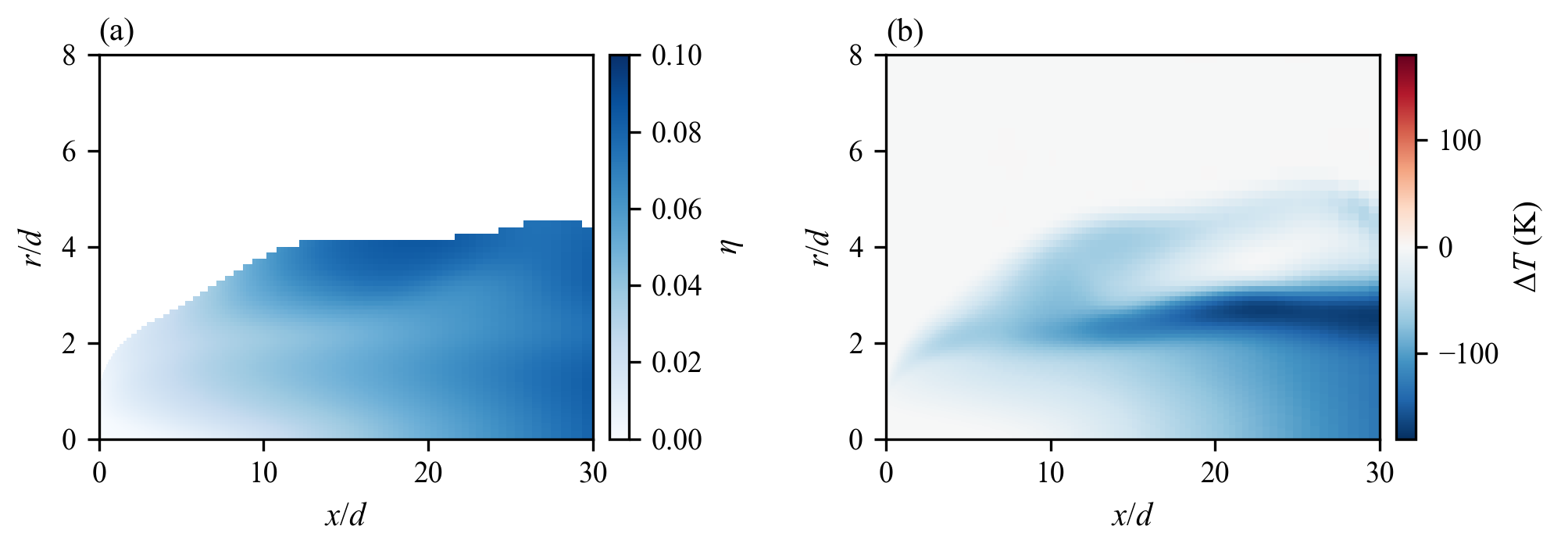}
\caption{Radiative heat loss at normal gravity (ULe, $30d\times10d$, $\Uco=0.15$\,m/s): (a) mean normalized
enthalpy defect $\eta$ of the radiating flame; (b) change of the mean temperature caused by radiation.}
\label{SI-fig:rad1g}
\alttext{Two color maps of the radiating flame at normal gravity: the normalized enthalpy defect, below 0.09 and growing downstream, and the change of the mean temperature caused by radiation, which reaches about $-170$\,K between $x/d=20$ and 30.}
\end{figure}

At normal gravity, the normalized defect stays below 0.09 but grows with the distance from the burner, and the cooling
reaches about 170\,K at $x/d=20$--30 (Fig.~\ref{SI-fig:rad1g}).

\FloatBarrier
%=====================================================================================================================
\section{Numerical method and tests}
\label{SI-sec:num}
The equations are discretized on a staggered grid and advanced with a second-order iterative scheme, implicit in the
radial and azimuthal terms and explicit in the axial terms. The pressure Poisson equation is solved by a Fourier
transform in $\theta$ and multigrid in $x$ and $r$, and the axial Courant number of 1.0 includes the convective and
viscous terms. Table~\ref{SI-tab:num} lists the tests at normal gravity, each against its reference.

At $0.05\,g_0$, the domain was widened to a radius of $45d$ and lengthened to $158d$. On the wider domain, the trailing
front of the radiating flame recedes 5\% faster; on the longer one, its position and amplitude agree with those on the
$94d$ domain to within $0.02d$ up to $x/d=88$. The adiabatic flame keeps a weak oscillation near the end of the $94d$
domain, with an onset at $x/d=77$--78 after 9.75\,s; continued on the $158d$ domain, its onset recedes from
$x/d=96$--103 to 123 within 5\,s while its swing decays by 37--42\% per 0.5\,s. On 16 azimuthal planes, the
reduced-gravity flames stay axisymmetric, also with white noise added to the jet inflow. Simulations on a grid with twice the radial resolution were started from base-grid states at $0.1\,g_0$. On this
grid, the oscillation is likewise sustained at $\chi=0.5$ and recedes at 0.75 and 1. At $\chi=1$, no swing above $0.25d$ remains within $x\le84d$ on either grid by the end of the record. The criteria of Section~II~E of the article then give an undecided state on the base grid
and a lost state on the refined grid (Table~\ref{SI-tab:dose}). Where the
co-flow has been changed, the state is read only from windows after two transit times of the co-flow, when the
co-flow near the lateral boundary is within 4\% of its final value upstream of the onset.

\begin{table}[!htbp]
\caption{Numerical tests at normal gravity ($\Uco=0.15$\,m/s; DD-$\alpha$ closure unless noted). p-p: median ratio of
the phase-averaged peak-to-peak temperature at the flickering measurement points to that of the reference.}
\label{SI-tab:num}
\footnotesize\setlength{\tabcolsep}{4pt}
\begin{tabular}{p{0.26\textwidth}p{0.13\textwidth}p{0.29\textwidth}p{0.23\textwidth}}
\toprule
\RR Test (reference) & \RR Frequency & \RR Flame position, mean and rms $T$ & \RR Amplitude \\
\midrule
\RR Courant number 0.5 / 1.0 / 2.0 (0.25) & \RR 13.92\,Hz in all & \RR $|\Delta T|\le0.1$ / 0.3 / 0.7\,K & \RR p-p ratio 0.9999 / 0.9995 / 0.9988 \\
\RR Axial and radial grid $\times2$ (base) & \RR 13.79 (13.72)\,Hz & \RR 1000\,K radius within $0.01d$, $x/d\le8$ & \RR edge swing $+1.1$ to $+4.8\%$ \\
\RR 8 and 16 azimuthal planes (4) & \RR 13.92\,Hz in all & \RR mean / rms $T$ within 0.4 / 1.1\,K & \RR p-p within 1.4\% \\
\RR Domain length $62d$ ($30d$) & \RR 13.89 (13.92)\,Hz & \RR mean / rms $T$ within 2.5 / 3.7\,K at $x/d\le16$ & \RR p-p within 2.1\% \\
\RR Radial extent $30d$ ($10d$), four closures & \RR $+0.4$ to $+2.4\%$ & \RR 1000\,K radius at $x/d=4$ within $0.03d$ & \RR swing at $x/d=8$ within $0.13d$ \\
\bottomrule
\end{tabular}
\alttext{Table of numerical tests at normal gravity (Courant number, grid refinement, azimuthal planes, domain length, and radial extent) and their small effects on the frequency, flame position, mean and rms temperature, and oscillation amplitude.}
\end{table}

\FloatBarrier
%=====================================================================================================================
\section{Local stability analysis}
\label{SI-sec:lsa}
Saddle points of the hot-layer branch are located by Newton iteration and retained only if they are pinch points by
the Briggs--Bers criterion; the trailing-edge velocity $V_-$ follows by continuation along rays from the absolute
saddle. The complete linearization of Section~II~F of the article also perturbs all transport terms, and its radiative source-response term is $-\varepsilon\rho\,(q_\xi\xi'+q_hh')$, with $\varepsilon=1$ for the
responding and $\varepsilon=0$ for the frozen source; the base sink acting on the density perturbation is kept in
both cases. Its coupling terms
change the stability far more than the radiative response: with the dilatation and the composition coupling, no absolutely
unstable region remains in the flickering flames at 0.165 and $0.25\,g_0$, and the absolute frequency at normal gravity
falls to 6.1--6.8\,Hz, against the simulated 14.5\,Hz. The simplified model of Eqs.~(2)--(4), which locates the onsets
and gives the frequency to within 20\% (Section~\ref{SI-sec:onset}), is therefore used for the absolute and convective
character of the flows, and the complete linearization only for the change caused by the radiative response.

\FloatBarrier
%=====================================================================================================================
\section{Measured temperature histories}
\label{SI-sec:cars}
\begin{figure}[!htbp]
\centering
\includegraphics[width=\textwidth]{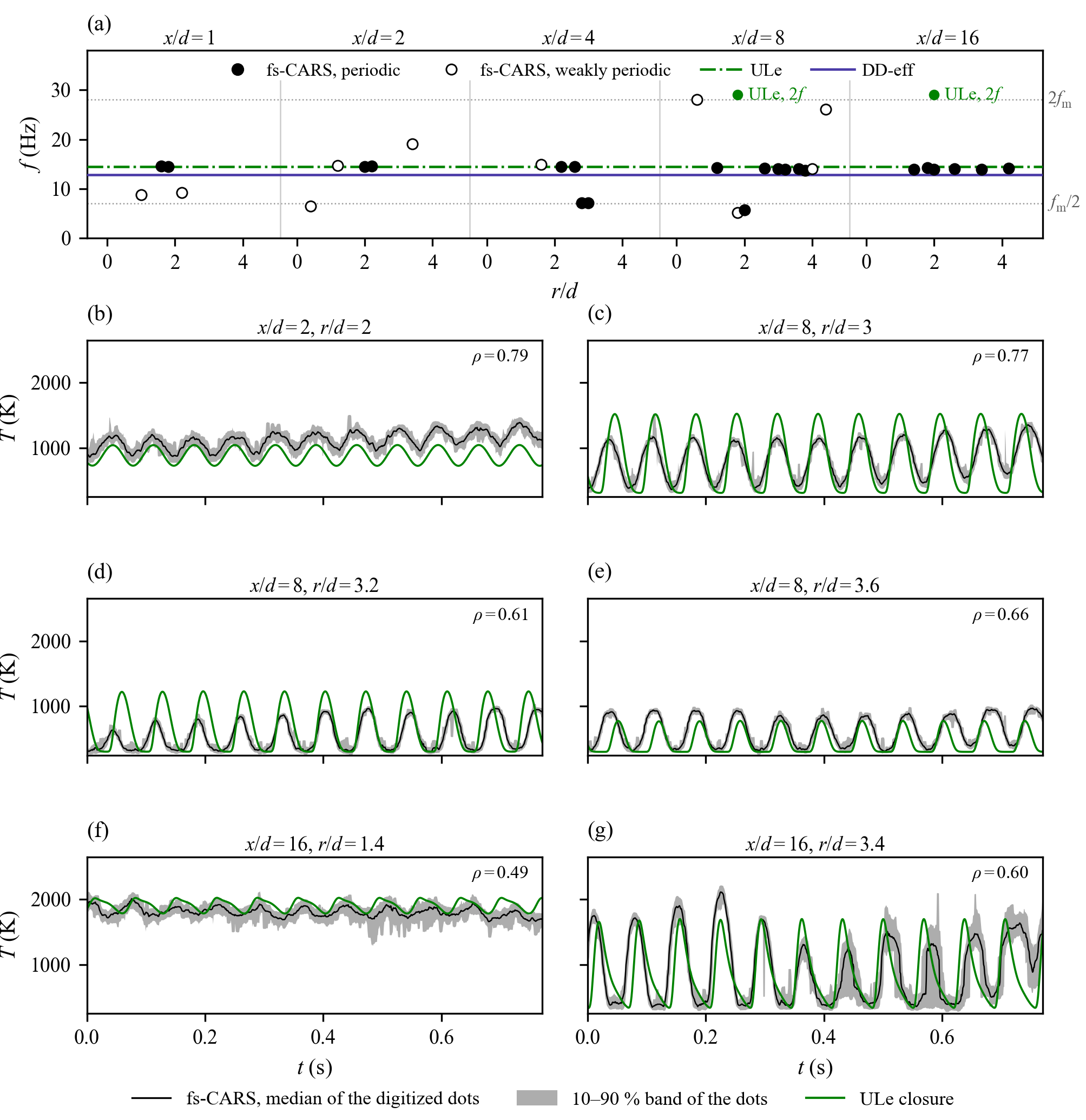}
\caption{(a) Dominant frequency of the temperature at every measurement point, grouped by height: fs-CARS (filled:
periodic; open: weakly periodic) and the simulations (lines; symbols: points at $2f$). (b--g) Temperature histories at
the six selected points: fs-CARS (median and 10--90\% band of the digitized dots) and the ULe simulation,
shifted to the best correlation $\rho$.}
\label{SI-fig:cars6}
\alttext{Top panel: dominant frequency of the temperature at every measurement point, grouped by height, near 14\,Hz at most periodic points in the measurement and the simulations. Six lower panels: measured and simulated temperature histories at selected points, with similar period and waveform.}
\end{figure}

Figure~\ref{SI-fig:cars6} compares the dominant frequency of the temperature at every measurement point with the
simulations, and the temperature histories at six points with the ULe simulation. The six points are those at which
the measured signal is periodic and the computed mean temperature lies within 250\,K of the measured one; at each, the
simulation is shifted in time to the best correlation, which reaches 0.49--0.79. At each measurement point, the dominant frequency is the peak of the Fourier spectrum of the 0.8\,s record; the
frequency quoted in the article is the median over the 18 points with a coherent oscillation. The measured swing of the flame edge (Fig.~3e of the article; Table~\ref{SI-tab:closures}) is inferred from
the measured mean temperature and its coherent fluctuation with a moving-edge model, which recovers the computed swing
to within 5\% when applied to the simulation.

\FloatBarrier
%=====================================================================================================================
\section{Wave structure and stability at normal gravity}
\label{SI-sec:wave}
\begin{figure}[!htbp]
\centering
\includegraphics[width=\textwidth]{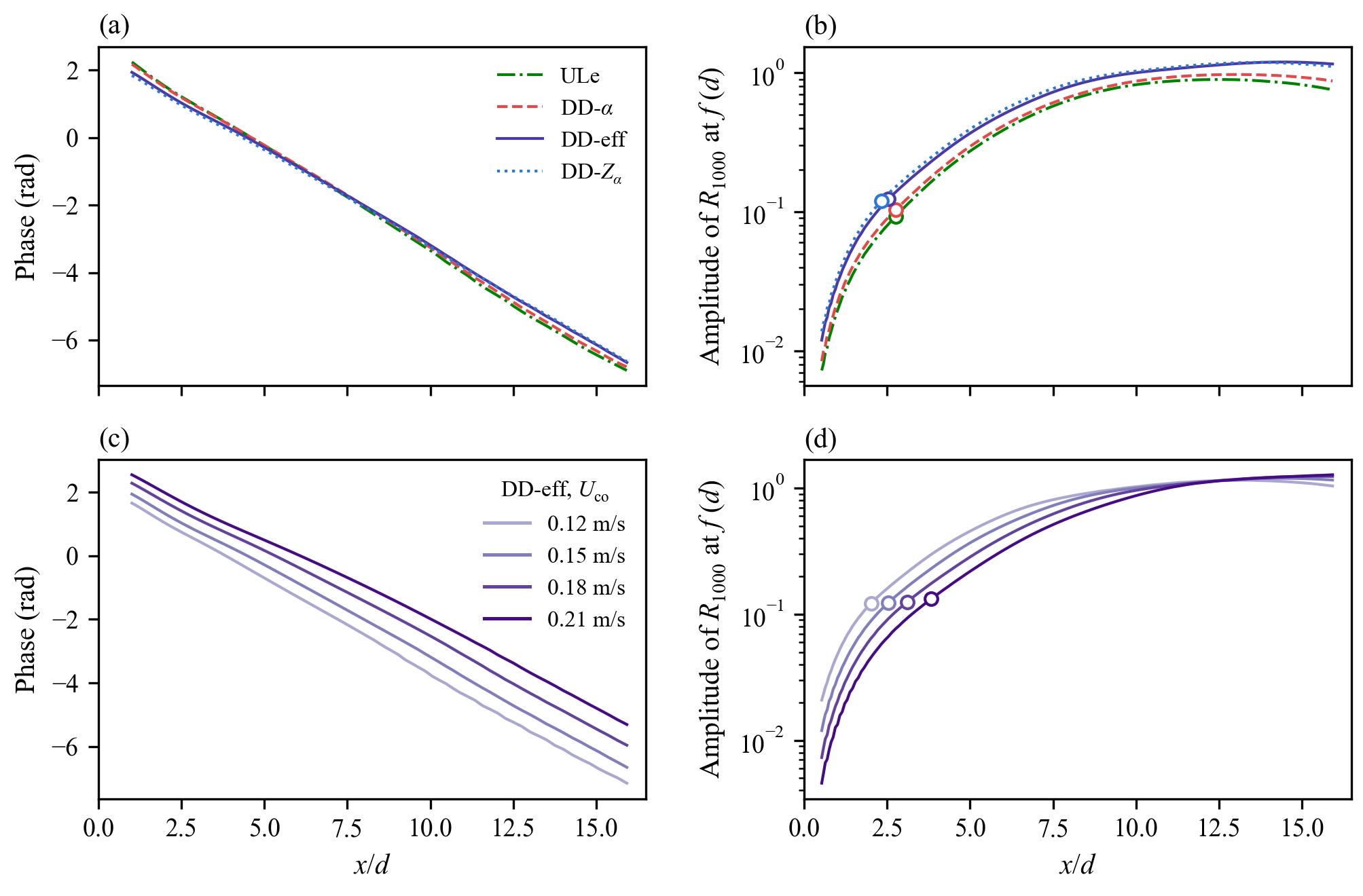}
\caption{Edge oscillation at the flicker frequency at normal gravity: phase (a, c) and amplitude (b, d) of the outer
1000\,K radius; (a, b) the four closures at $\Uco=0.15$\,m/s; (c, d) DD-eff with $\Uco=0.12$--0.21\,m/s. Open circles:
amplitude at 10\% of its maximum.}
\label{SI-fig:wave}
\alttext{Four line plots of the phase and amplitude of the flame-edge oscillation against the axial distance at normal gravity, for four closures and for several co-flow velocities; the phase falls linearly, and the amplitude grows exponentially before it saturates.}
\end{figure}

\begin{figure}[!htbp]
\centering
\includegraphics[width=\textwidth]{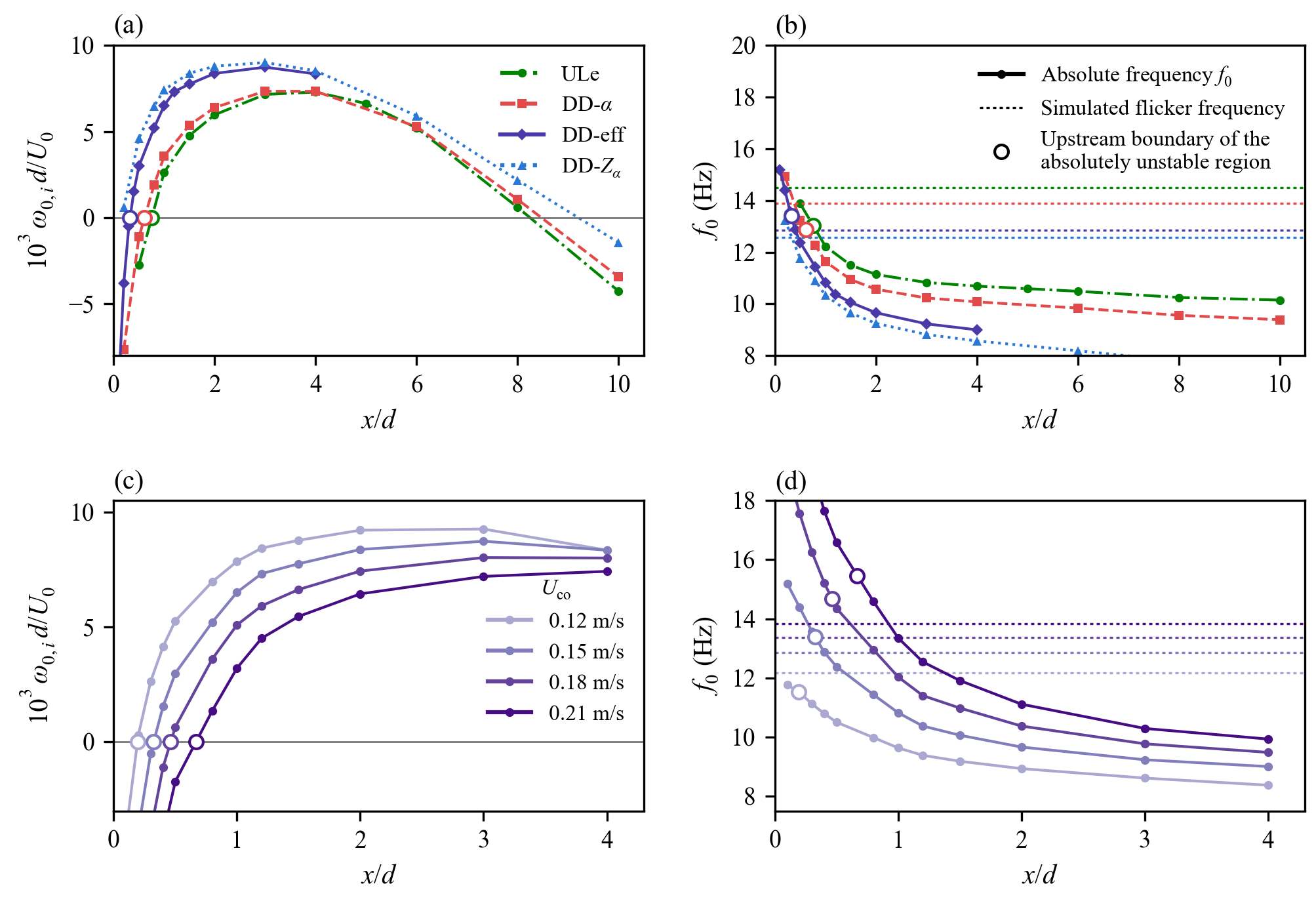}
\caption{Absolute growth rate and absolute frequency of the axisymmetric mode of the hot layer at normal gravity:
(a, b) the closures at $\Uco=0.15$\,m/s; (c, d) DD-eff with $\Uco=0.12$--0.21\,m/s. Open circles: upstream boundary of
the absolutely unstable region.}
\label{SI-fig:stab1g}
\alttext{Four line plots of the absolute growth rate and the absolute frequency of the hot-layer mode against the axial distance at normal gravity, for four closures and for several co-flow velocities; every closure shows an absolutely unstable region near the burner.}
\end{figure}

At normal gravity, the edge oscillation is a single traveling wave: its phase falls linearly with $x$, and its amplitude
grows exponentially from the burner, at a spatial rate of 0.60--0.76 per diameter, before it saturates
(Fig.~\ref{SI-fig:wave}). The wavelength scales with the flame radius, $\lambda\approx4.05R_4$, and the wider DD-eff flame
oscillates more slowly because its wave is both longer and slower. For every closure, the local analysis finds an
absolutely unstable region near the burner (Fig.~\ref{SI-fig:stab1g}).

\FloatBarrier
%=====================================================================================================================
\section{Emission factor and perturbations at reduced gravity}
\label{SI-sec:dose}
\begin{figure}[!htbp]
\centering
\includegraphics[width=\textwidth]{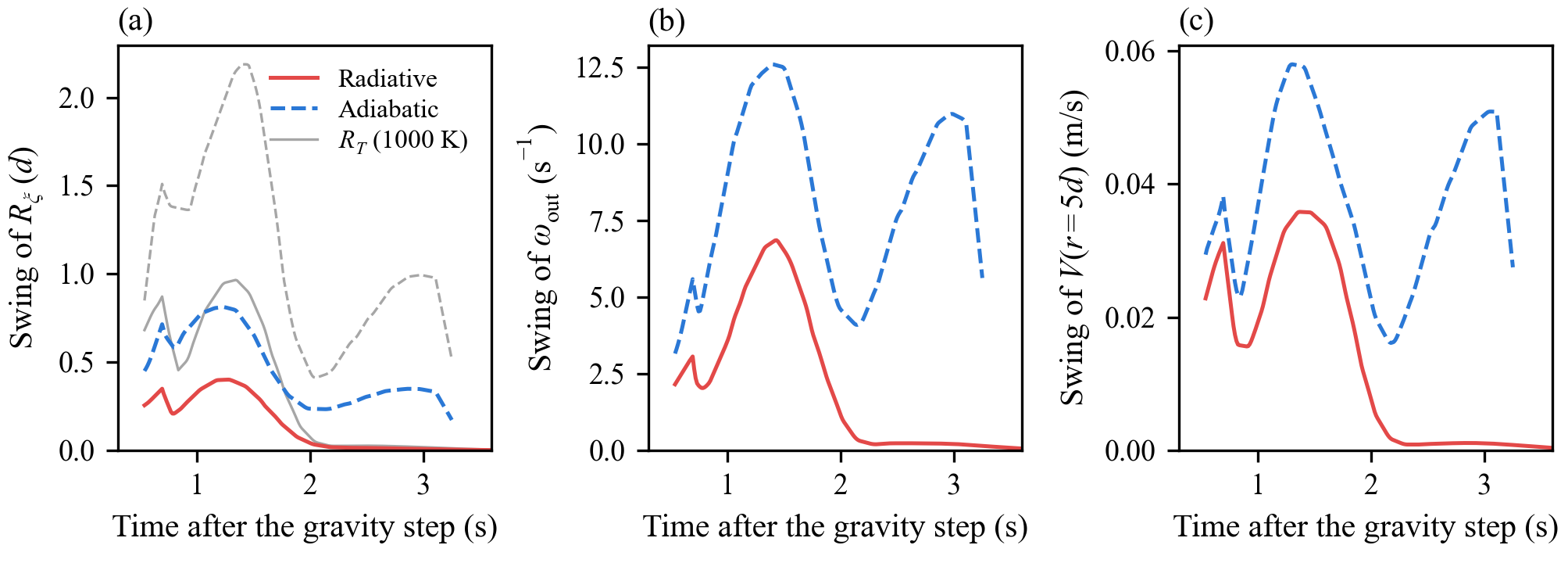}
\caption{Perturbations at $x/d=60$ of the radiating and the adiabatic flame after the step to $0.05\,g_0$:
(a) stoichiometric surface (thin gray: 1000\,K isotherm); (b) azimuthal vorticity outside the radius of peak
temperature; (c) radial velocity at $r=5d$.}
\label{SI-fig:pert}
\alttext{Three time histories at $x/d=60$ after the step to $0.05\,g_0$ comparing the radiating and the adiabatic flame: stoichiometric surface, outer-layer vorticity, and radial velocity, all of which oscillate more weakly with radiation.}
\end{figure}

Table~\ref{SI-tab:dose} lists the simulations with the emission factor $\chi$ (Section~III~B of the article). At
$0.1\,g_0$, the emission is switched on in the adiabatic flame 6.7\,s after the gravity step ($\chi>0$) and off in the
radiating flame 9.75\,s after it ($1\to0$); at $0.165\,g_0$, it is raised in the radiating flame. Figure~\ref{SI-fig:pert}
compares the radiating and the adiabatic flame at $x/d=60$ after the step to $0.05\,g_0$, in quantities that
radiation changes only through the flow; the ratio of 1.6--2.2 given in Section~III~B of the article is read between
0.8 and 2\,s after the step.

\begin{table}[!htbp]
\caption{Oscillation against the emission factor $\chi$ ($\Uco=0.15$\,m/s): state, onset $x_\mathrm{on}$ and frequency
$f$ in the final record, and recession speed of the onset (in parentheses: grid with twice the radial resolution,
continued from a state of the base grid).}
\label{SI-tab:dose}
\footnotesize
\begin{tabular}{ccccc}
\toprule
$\chi$ & State & $x_\mathrm{on}$ ($d$) & $f$ (Hz) & Recession ($d$/s) \\
\midrule
\multicolumn{5}{l}{$0.1\,g_0$} \\
0 & sustained & 20.0 & 3.95 & 0 \\
0.5 & sustained & 39.7 (37.7) & 4.67 (4.53) & 0 (0) \\
0.7 & receding & 49.0 & 4.53 & 2.0 \\
0.75 & receding (receding) & 52--54 (61--64) & 4.67 (5.50) & 2.7 (3.4) \\
0.8 & receding & 56--58 & 4.67 & 3.7 \\
1$^\mathrm{b)}$ & undecided$^\mathrm{a)}$ (lost) & -- & 4.97 (5.14) & $\approx9$ ($\approx10$) \\
$1\to0$ & sustained & 20.0 & 3.95 & 0 \\
\midrule
\multicolumn{5}{l}{$0.165\,g_0$} \\
1 & sustained & 15.1 & 5.50 & 0 \\
1.15 & sustained & 15.6--15.9 & 5.50 & 0 \\
1.65 & receding & 26.0 & 5.70 & 2.5 \\
\bottomrule
\end{tabular}
\par\smallskip{\footnotesize $^\mathrm{a)}$Receded beyond $x/d=84$: from 5.9\,s on no swing above $0.25d$ remains within
$x\le84d$. $^\mathrm{b)}$Emission switched on in the oscillating adiabatic flame; the radiating flame after the
gravity step, which has lost its flicker, is listed in Table~II of the article.}
\alttext{Table of the emission-factor series at 0.1 and $0.165\,g_0$: state, onset, frequency, and recession speed for each emission factor, with refined-grid results in parentheses; at $0.1\,g_0$ the oscillation is sustained at $\chi=0.5$ and recedes from $\chi=0.7$ on.}
\end{table}

\FloatBarrier
%=====================================================================================================================
\section{Gravity pulses and radiative damping}
\label{SI-sec:pulse}
Each pulsed flame is compared with the same flame without the pulse, computed on the same grid partition, so that the
difference is the response to the pulse. At $0.1\,g_0$, the oscillation of the frozen-source flame meets the criterion
of a sustained oscillation in the last windows of its 12.1\,s record (largest swing $+0\%$ over the last second, onset
$+0.7d$ over the last 2\,s). Over the last 6\,s, however, its onset recedes from $x/d=42.3$ to 45.7, and its swing
falls from 1.5 to $0.9d$ at $x/d=60$ and from 4.5 to $3.6d$ at its largest, at $x/d=83.7$. Without the pulse, the
flame stays steady, with swings of $0.064d$ or less. Upstream of the onset, this flame keeps the hot layer of the radiating flame: density,
velocity jump, and radius at the temperature peak agree to within 0.06\%.

In the complete linearization, the radiative response lowers the absolute growth rate by 0.34--0.55\,s$^{-1}$ in six
flames between $g_0$ and $0.05\,g_0$. Its local contributions partly cancel because they differ in phase, leaving
32--49\% of their summed magnitude. Table~\ref{SI-tab:vmeps} lists the upstream spreading speed on fixed flows with the
response off and on, together with its first-order change, Eq.~(8) of the article.

\begin{table}[!htbp]
\caption{Upstream spreading speed $S=\Uco-V_-$ with the radiative response off ($\varepsilon=0$) and on
($\varepsilon=1$), from the complete linearization on fixed time-averaged flows of the radiating flames
($\Uco=0.15$\,m/s; at $0.25\,g_0$, 0.25\,m/s); first order: $\mathrm{Im}[\partial_\varepsilon\omega(k_-)]/|\mathrm{Im}\,k_-|$.}
\label{SI-tab:vmeps}
\footnotesize
\begin{tabular}{ccccccc}
\toprule
$g/g_0$, $x/d$ & $S(0)$ (m/s) & $S(1)$ (m/s) & $S(1)-S(0)$ (m/s) & first order (m/s) &
$\mathrm{Im}\,\partial_\varepsilon\omega(k_-)$ (s$^{-1}$) & $\mathrm{Im}\,k_-$ ($1/d$) \\
\midrule
0.25, 25 & 0.1940 & 0.1849 & $-0.0091$ & $-0.0093$ & $-0.41$ & $-0.218$ \\
0.165, 15 & 0.1179 & 0.1083 & $-0.0096$ & $-0.0100$ & $-0.45$ & $-0.223$ \\
0.1, 30 & 0.1069 & 0.0967 & $-0.0102$ & $-0.0103$ & $-0.42$ & $-0.201$ \\
0.05, 20 & 0.0388 & 0.0267 & $-0.0121$ & $-0.0128$ & $-0.51$ & $-0.199$ \\
0.05, 40 & 0.0464 & 0.0345 & $-0.0120$ & $-0.0119$ & $-0.45$ & $-0.187$ \\
\bottomrule
\end{tabular}
\alttext{Table of the upstream spreading speed with the radiative response off and on at five stations of four radiating flames, its decrease of about 0.01\,m/s, and the first-order estimate, which agrees to within 5\%.}
\end{table}

\FloatBarrier
%=====================================================================================================================
\section{Co-flow, onset and frequency}
\label{SI-sec:onset}
\begin{figure}[!htbp]
\centering
\includegraphics[width=\textwidth]{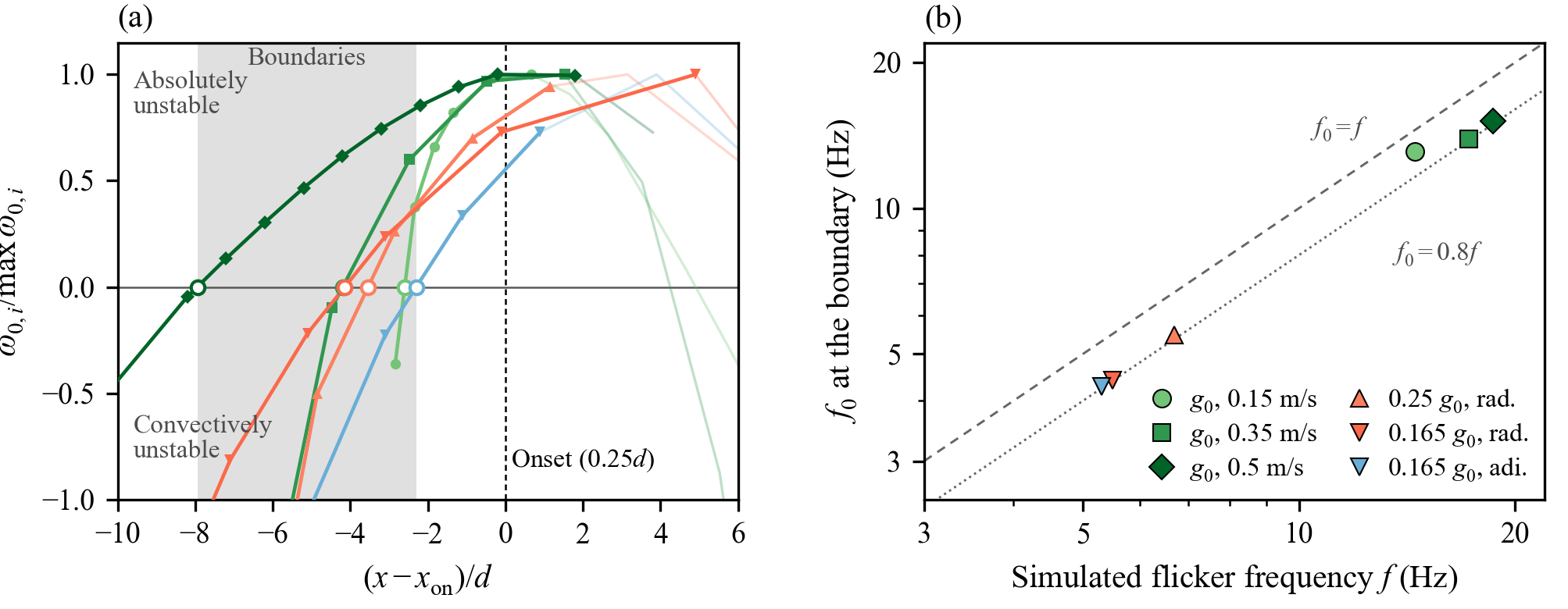}
\caption{Local stability of six of the ten oscillating flames with an absolutely unstable region (symbols as in Fig.~11c of
the article): (a) absolute growth rate, divided by its largest value along the flame, against the distance from the
onset (faint: downstream of the onset; circles: upstream boundaries; gray band: their range); (b) absolute frequency at
the upstream boundary against the simulated flicker frequency.}
\label{SI-fig:lsaextra}
\alttext{Left: normalized absolute growth rate against the distance from the onset for six flames, changing sign a few diameters upstream of the onset. Right: absolute frequency at the upstream boundary against the simulated flicker frequency, 10--20\% below it.}
\end{figure}

\begin{figure}[!htbp]
\centering
\includegraphics[width=3.37in]{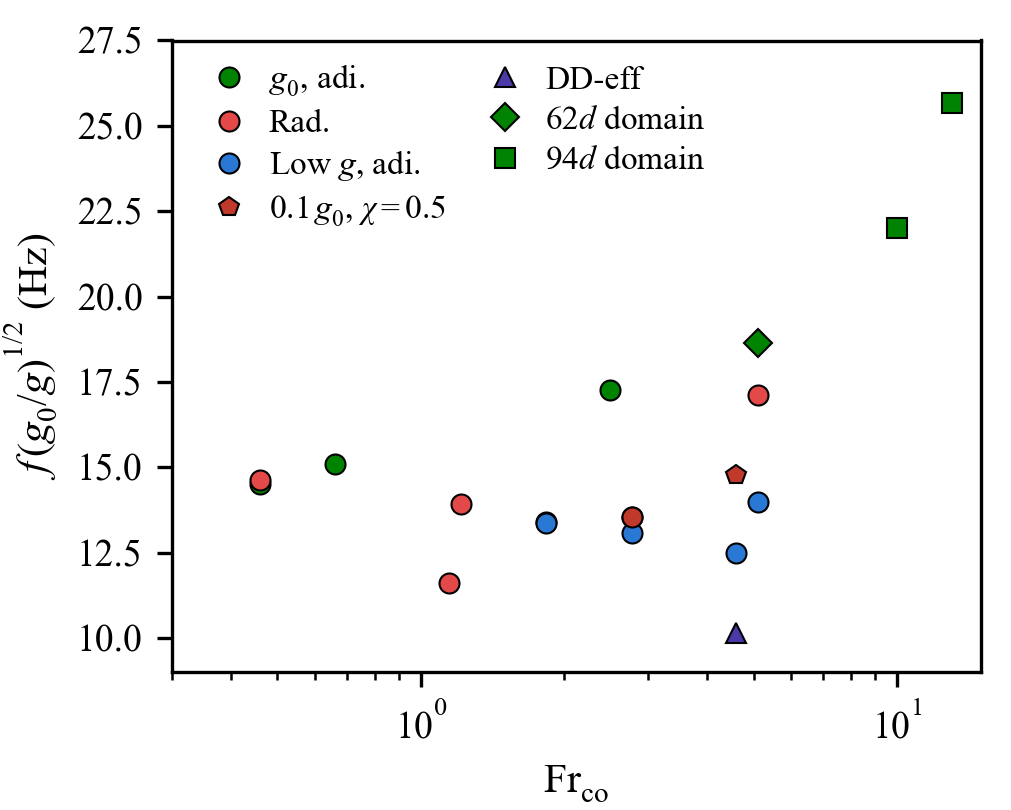}
\caption{Frequency of the sustained oscillations scaled by $(g_0/g)^{1/2}$ against $\Frco$.}
\label{SI-fig:fscaled}
\alttext{Scatter plot of the flicker frequency scaled by the square root of gravity against the co-flow Froude number for all sustained cases; the points do not collapse on one curve and range from about 10 to 26\,Hz.}
\end{figure}

Table~\ref{SI-tab:coflow} lists the oscillation at normal gravity against the co-flow velocity (Fig.~11a of the article).
Figure~\ref{SI-fig:lsaextra} shows the absolute growth rate along six of the ten flames with an absolutely unstable
region, at normal gravity with three co-flows and at 0.165 and $0.25\,g_0$, and the absolute frequency at the
upstream boundary of this region; measured from the onset defined with $0.25d$, the boundary lies 2.3--7.9$d$
upstream. Scaled by $(g_0/g)^{1/2}$, the frequency of the sustained oscillations is not ordered by $\Frco$, unlike the
onset (Fig.~\ref{SI-fig:fscaled}): at $\Frco=4.6$--5.1 and with unity Lewis numbers, it ranges from 12.5\,Hz at $0.1\,g_0$
to 18.6\,Hz at normal gravity.

\begin{table}[!htbp]
\caption{Oscillation at normal gravity against the co-flow velocity (ULe): frequency $f$, flame radius $R_4$, swing
of the flame edge at three heights (Section~II~E of the article, last 0.5\,s), and onset $x_\mathrm{on}$.}
\label{SI-tab:coflow}
\footnotesize
\begin{tabular}{ccccccccc}
\toprule
$\Uco$ (m/s) & $\Frco$ & Domain & $f$ (Hz) & $R_4$ ($d$) & \multicolumn{3}{c}{Swing ($d$) at $x/d=$} & $x_\mathrm{on}$ ($d$) \\
 & & & & & 4 & 8 & 16 & \\
\midrule
0.15 & 0.46 & $30d\times10d$ & 14.51 & 2.31 & 0.36 & 1.28 & 1.58 & 3.3 \\
0.18 & 0.66 & $30d\times10d$ & 15.09 & 2.29 & 0.26 & 1.07 & 1.80 & 4.0 \\
0.35 & 2.50 & $30d\times10d$ & 17.26 & 2.17 & 0.05 & 0.23 & 1.77 & 8.3 \\
0.50 & 5.10 & $30d\times10d$ & 18.64 & 2.06 & 0.00 & 0.02 & 0.15 & 18.0 \\
0.50 & 5.10 & $62d\times10d$ & 18.63 & -- & -- & -- & -- & 22.2 \\
0.70 & 9.99 & $94d\times30d$ & 22.0 & -- & 0.00 & 0.01 & 0.02 & 27.2 \\
0.80 & 13.05 & $94d\times30d$ & 25.7 & -- & 0.00 & 0.00 & $\le0.01$ & 36.0 \\
\bottomrule
\end{tabular}
\alttext{Table of the oscillation at normal gravity for co-flow velocities from 0.15 to 0.8\,m/s: Froude number, domain, frequency, flame radius, edge swing at three heights, and onset, which moves downstream as the co-flow increases.}
\end{table}

\FloatBarrier
%=====================================================================================================================
\section{Calibration of the limit relation}
\label{SI-sec:calib}
The constants of Eq.~(9) of the article are calibrated on the 27 unity-Lewis-number cases computed before the test at
$0.165\,g_0$: 16 sustained, which should lie at $U_\mathrm{c}/\Uco\ge1$, and 7 lost and 4 receding, which should lie
at $U_\mathrm{c}/\Uco\le1$. A parameter pair $(c,b)$ satisfies a tolerance $\tau$ when every sustained case lies at $U_\mathrm{c}/\Uco\ge1-\tau$ and every lost or receding case at
$U_\mathrm{c}/\Uco\le1+\tau$. The selected constants, $c=3.05$ and $b=0.09$\,m/s, place the sustained cases at
$U_\mathrm{c}/\Uco\ge0.965$ and the others at $U_\mathrm{c}/\Uco\le1.007$, so that $\tau=3.5\%$; the band of
Fig.~12b of the article marks the interval 0.965--1.007 in which the two groups overlap. Every pair that satisfies the same tolerance has $c=3.05$--3.13 and $b=0.084$--0.106\,m/s. The receding flames constrain $b$:
without them, the pairs within this tolerance reach $b=0.059$--0.106\,m/s.

\clearpage
%=====================================================================================================================
\section{Symbols and abbreviations}
\label{SI-sec:symbols}

\begin{table}[!htbp]
\caption{Symbols and abbreviations.}
\label{SI-tab:symbols}
\footnotesize\setlength{\tabcolsep}{4pt}
\begin{tabular}{p{0.12\textwidth}p{0.34\textwidth}p{0.13\textwidth}p{0.32\textwidth}}
\toprule
\RR Symbol & \RR Meaning & \RR Symbol & \RR Meaning \\
\midrule
\RR $d$ & \RR inner diameter of the fuel tube, 5\,mm & \RR $\beta$ & \RR buoyancy factor $\rho_\mathrm{air}/\rho_\mathrm{hot}-1$ at the radius of peak temperature \\
\RR $U_0$ & \RR bulk velocity of the fuel jet, 7.7\,m/s & \RR $\gamma$ & \RR rate of the radiative loss of buoyancy, Eq.~(5) \\
\RR $\Uco$ & \RR co-flow velocity & \RR $t_\mathrm{res}$ & \RR residence time of the gas in the hot layer \\
\RR $g$, $g_0$ & \RR gravitational acceleration; normal gravity & \RR $U_\mathrm{hot}$ & \RR axial velocity at the radius of peak temperature \\
\RR $\mathrm{Re}$ & \RR Reynolds number $U_0d/\nu_\mathrm{F}$ & \RR $\Delta U$ & \RR velocity difference between the radius of peak temperature and the co-flow at $r=12d$ \\
\RR $\Frco$ & \RR co-flow Froude number $\Uco^2/(gd)$ & \RR $\omega$, $k$ & \RR angular frequency and axial wavenumber of a perturbation \\
\RR $x$, $r$ & \RR axial and radial coordinates & \RR $\omega_0$, $\omega_{0,i}$ & \RR absolute frequency and absolute growth rate \\
\RR $\xi$, $\xi_\mathrm{st}$ & \RR mixture fraction; its stoichiometric value & \RR $x_\mathrm{a}$ & \RR upstream boundary of the absolutely unstable region \\
\RR $\Delta h$, $\eta$ & \RR enthalpy defect; normalized defect & \RR $V_-$, $k_-$ & \RR velocity of the trailing edge of the impulse response; its saddle point \\
\RR $q_\mathrm{rad}$ & \RR radiative sink per unit mass & \RR $S$ & \RR upstream spreading speed relative to the co-flow, $\Uco-V_-$ \\
\RR $\chi$ & \RR emission factor: emission relative to the optically thin model & \RR $\zeta$ & \RR radial displacement of the base fields by a perturbation \\
\RR $\Omega_\mathrm{r}$ & \RR radiative cooling rate of the products, $\max(q_\mathrm{rad}/c_pT)$ & \RR $U_\mathrm{c}$ & \RR critical co-flow velocity, the largest at which the flicker persists, Eq.~(9) \\
\RR $R(x,t)$ & \RR outer flame edge: largest radius of the 1000\,K isotherm & \RR $c$, $b$ & \RR buoyant and radiative constants of Eq.~(9) \\
\RR swing & \RR peak-to-peak value of $R$ about its running mean, in windows of 1.25 periods & \RR $\Pi$, $B$ & \RR radiation number $\Omega_\mathrm{r}(d/g)^{1/2}$; $b/(\Omega_\mathrm{r}d)$ \\
\RR $x_\mathrm{on}$ & \RR onset: most upstream position with a swing above $0.25d$ & \RR $\kappa$ & \RR diffusivity of the density perturbation, $\mu/(\rho\,\mathrm{Pr})$ \\
\RR $f$, $R_4$ & \RR flicker frequency; outer 1000\,K radius of the mean flame at $x/d=4$ & \RR $\varepsilon$ & \RR strength of the radiative response in the linear model \\
\RR $\lambda$ & \RR wavelength of the edge oscillation & & \\
\midrule
\RR ULe & \RR unity-Lewis-number closure & \RR DD-eff & \RR differential-diffusion closure with the effective diffusivity of $\xi_\mathrm{B}$ \\
\RR fs-CARS & \RR femtosecond coherent anti-Stokes Raman scattering & \RR DD-$\alpha$, DD-Z$\alpha$ & \RR further differential-diffusion closures (Section~\ref{SI-sec:closures}) \\
\bottomrule
\end{tabular}
\alttext{Table listing, in two pairs of columns, the symbols and abbreviations used in the article and the supplementary material, each with its meaning and, where one exists, its defining equation.}
\end{table}

\end{document}